\documentclass[12pt]{article}
\usepackage[totalwidth=460truept,totalheight=628truept]{geometry}
\usepackage{latexsym,graphicx,amsfonts,amssymb,amsmath,xcolor,changepage,bm}
\usepackage[hypertexnames=false,hidelinks]{hyperref}

\def\theequation{\arabic{section}.\arabic{equation}}

\renewcommand{\theequation}{\thesection.\arabic{equation}}
\global\arraycolsep=1truept
\renewcommand{\theequation}{\arabic{section}.\arabic{equation}}
\null

\begin{document}

\qquad \medskip

\begin{center}
{\Huge \textbf{\negthinspace Gauge Theories and Quantum Gravity\negthinspace 
}}

\vskip.6truecm

{\Huge \textbf{in a Finite Interval of Time,}}

\vskip.65truecm

{\Huge \textbf{on a Compact Space Manifold}}

\vskip 1truecm

\textsl{Damiano Anselmi}

\vskip.1truecm

{\small \textit{Dipartimento di Fisica \textquotedblleft
E.Fermi\textquotedblright , Universit\`{a} di Pisa, Largo B.Pontecorvo 3,
56127 Pisa, Italy}}

{\small \textit{INFN, Sezione di Pisa, Largo B. Pontecorvo 3, 56127 Pisa,
Italy}}

{\small damiano.anselmi@unipi.it}

\vskip 1.5truecm

\textbf{Abstract}
\end{center}

We study gauge theories and quantum gravity diagrammatically in a finite
interval of time $\tau $, on a compact space manifold $\Omega $. The
initial, final and boundary conditions are formulated in gauge invariant and
general covariant ways by means of purely virtual extensions of the
theories, which allow us to \textquotedblleft trivialize\textquotedblright\
the local symmetries and switch to invariant fields (the invariant metric
tensor, invariant quark and gluon fields, etc.). The evolution operator $%
U(t_{\text{f}},t_{\text{i}})$ is worked out for arbitrary initial and final
states, as well as general boundary conditions on $\partial \Omega $. We
show that $U(t_{\text{f}},t_{\text{i}})$ is well defined and
diagrammatically unitary for every $\tau =t_{\text{f}}-t_{\text{i}}<\infty $%
. The formulation is extended to include purely virtual particles. In
quantum gravity, where the cosmological constant $\Lambda _{C}$ challenges
the definition of an $S$ matrix, the results allow us to prove unitarity at $%
\tau <\infty $. We work out the frequencies and eigenfunctions in some
explicit examples, including Yang-Mills theory on the finite cylinder.

\vfill\eject

\section{Introduction}

\label{intro}\setcounter{equation}{0}

Perturbative quantum field theory mainly focuses on the calculation of $S$
matrix amplitudes, which describe scattering processes among asymptotic
states, where the incoming and outgoing particles are separated by an
infinite amount of time. This approximation is good for most practical
purposes, especially in collider physics. However, it is just an
approximation. From a theoretical point of view, it does not provide a
completely satisfactory understanding. A more powerful and general approach
is required, where the key issues (such as locality, renormalizability and
unitarity, among the main ones, and then symmetries, anomalies, the anomaly
cancellation, etc.) are understood without making this simplification.

It is possible \cite{MQ} to formulate perturbative quantum field theory
diagrammatically in a finite interval of time $\tau =t_{\text{f}}-t_{\text{i}%
}$, and on a compact space manifold $\Omega $, so as to move all the details
about the restrictions to finite $\tau $ and compact $\Omega $ away from the
internal sectors of the diagrams (apart from the discretizations of the loop
momenta), and code them into external sources. The usual diagrammatic
properties apply, or can be generalized with little effort. This way, the
evolution operator $U(t_{\text{f}},t_{\text{i}})$ can be calculated
perturbatively between arbitrary initial and final states, with arbitrary
boundary conditions on $\partial \Omega $. Unitarity, that is to say, the
equality $U^{\dag }(t_{\text{f}},t_{\text{i}})U(t_{\text{f}},t_{\text{i}})=1$%
, can be studied diagrammatically by means of the spectral optical
identities \cite{diagrammarMio}. The theory is renormalizable whenever it is
so at $\tau =\infty $, $\Omega =\mathbb{R}^{D-1}$, where $D$ denotes the
spacetime dimension. Purely virtual particles are introduced by removing the
on-shell contributions of some physical particles, and all the ghosts, from
the core diagrams, as explained in \cite{diagrammarMio}, and trivializing
their initial and final conditions.

In this paper we consider the cases of gauge theories and gravity in detail,
because certain issues that are specific to local symmetries deserve
attention, when $\tau $ is finite and the space manifold $\Omega $ is
compact. For example, we must specify the initial, final and boundary
conditions without breaking the local symmetries. We cannot just use the
gauge potential $A_{\mu }^{a}$ and the metric tensor $g_{\mu \nu }$, for
this purpose. Nor can we use the field strength $F_{\mu \nu }^{a}$, and the
curvature tensors $R$, $R_{\mu \nu }$, $R_{\mu \nu \rho \mathbb{\sigma }}$,
because they are not invariant.

What comes to the rescue is the purely virtual extension of gauge theories
and gravity formulated in ref.s \cite{AbsoPhys,AbsoPhysGrav}, which is based
on the introduction of extra bosonic fields, together with their
anticommuting partners. The extra fields can be used to perturbatively
\textquotedblleft dress\textquotedblright\ the non invariant fields and make
them invariant: we can build invariant gauge fields $A_{\text{d}}^{\mu }$,
invariant quark fields $\psi _{\text{d}}$, and an invariant metric tensor $%
g_{\mu \nu \text{d}}$. The ordinary physical quantities, such as the $S$
matrix amplitudes and the correlation functions of the usual (nonlinear)
invariant composite fields (like $F_{\mu \nu }^{a}F^{\mu \nu a}$, $\bar{\psi}%
\psi $, etc.) are unaffected. In addition, new, physical correlation
functions can be defined, and calculated perturbatively, such as those of
the invariant fields $A_{\text{d}}^{\mu }$, $\psi _{\text{d}}$ and $g_{\mu
\nu \text{d}}$. The reason why the extra fields must be purely virtual is to
preserve unitarity: if they were not purely virtual, the extended theory
would propagate ghosts.

These tools are also useful to provide invariant initial, final and boundary
conditions in gauge theories and gravity in a finite interval of time $\tau $%
, on a compact space manifold $\Omega $. A crucial simplification comes from
the possibility of \textquotedblleft trivializing\textquotedblright\ the
local symmetries, that is to say, reduce them to simple shifts of fields.
This is achieved by switching to the invariant variables $\psi _{\text{d}}$, 
$A_{\text{d}}^{\mu }$ and $g_{\mu \nu \text{d}}$, by means of a field
redefinition. Then, it is relatively straightforward to organize the action
efficiently, work out the eigenfunctions and the frequencies for the
expansions of the fields, and introduce coherent states \cite{cohe}, which
are crucial to study the $U(t_{\text{f}},t_{\text{i}})$ diagrammatics and
make perturbative calculations without introducing unnecessary burdens \cite%
{MQ}. The functional integral is \textit{defined} as the integral on the
coefficients of the expansions. The local symmetries are under control in
all the operations we make, so $U(t_{\text{f}},t_{\text{i}})$ is gauge
invariant and invariant under general coordinate transformations.

We illustrate the basic properties of our formalism in Yang-Mills theory on
two relatively simple space manifolds: the semi-infinite cylinder and the
finite cylinder.

The coherent states are the eigenstates of the annihilation operator. In the
Lagrangian approach, which we adopt here, the switch to coherent states is
just a change of variables in the functional integral, combined with a wise
way of setting the initial and final conditions. In quantum mechanics, we
switch from coordinates $q$ and momenta $p$ to $z\sim q+ip$, $\bar{z}\sim
q-ip$, and set the initial conditions on $z$, the final conditions on $\bar{z%
}$. In quantum field theory analogous operations are made on the fields.
Uncovering the specifics of these operations in gauge theories and gravity
is part of the problem we need to face, and its solution is given in the
paper. For convenience, we keep referring to the new variables by means the
Hamiltonian terminology \textquotedblleft coherent states\textquotedblright\ 
\footnote{%
Details on the correspondence between the operatorial approach to coherent
states and the functional integral can be found in the paragraph 9-1-2 of 
\cite{Itzykson-Zuber}.}.

\bigskip

Purely virtual particles are a key ingredient of the whole formulation, so
we discuss this concept in some detail. A theory that contains purely
virtual particles is built from an extended (possibly unphysical) theory%
\footnote{%
The extended theory is unphysical if it contains ghosts (fields with kinetic
terms multiplied by the wrong signs).}, which is quantized as usual (that is
to say, by means of the common diagrammatics, defined by the Feynman $%
i\epsilon $ prescription), and performing a certain set of operations on it,
like rearranging the diagrammatics, and making a projection on the space of
states, to define the physical space. The projection defines the final,
physical theory.

The new diagrammatics is built by removing the on-shell contributions of all
the ghosts $\chi _{\text{gh}}$, and possibly some physical particles $\chi _{%
\text{ph}}$, from the diagrams of the extended theory, at every order of the
perturbative expansion. This is done in one of the following equivalent
ways: $i$) a certain nonanalytic Wick rotation \cite{Piva,LWgrav}, $ii$)
dropping the spectral optical identities associated with the unwanted
on-shell contributions \cite{diagrammarMio} from the Cutkosky-Veltman
identities \cite{cutkoskyveltman,thooft} (which are the diagrammatic
versions of the unitarity equation $S^{\dag }S=1$), or $iii$)\ replacing the
standard diagrams with suitable combinations of non-time-ordered diagrams,
as shown in ref. \cite{PVP20}.

In addition, one has to make the projection mentioned above. At $\tau
=\infty $, the projection amounts to ignore the diagrams that have $\chi _{%
\text{gh}}$ and $\chi _{\text{ph}}$ on the external legs. When $\tau <\infty 
$, it amounts to choose trivial initial and final conditions for the
coherent states of $\chi _{\text{gh}}$ and $\chi _{\text{ph}}$. The final
theory is unitary, provided all the ghosts of the extended theory are
rendered purely virtual.

Certain aspects of the construction of theories with purely virtual
particles resemble what we normally do to gauge-fix a gauge theory, where we
extend the theory by including unphysical excitations, such as the
Faddeev-Popov ghosts, and project the extension away at the end. The crucial
difference is that, in the case of purely virtual particles, no symmetry is
there to help us. This is why we need to switch to a different
diagrammatics, before making the projection.

It is worth to stress that, before the projection, the extended theory is
just a mathematical tool to get to the correct, final theory. It is not
possible to solve the problem of ghosts by just changing the viewpoint on a
theory, or focusing on different quantities (e.g., \textquotedblleft
in-in\textquotedblright\ correlation functions, instead of \textquotedblleft
in-out\textquotedblright\ ones, or different prescriptions for the
propagators, such as the retarded potentials, instead of the Feynman one,
and so on), or moving back and forth among negative norms, unbounded
Hamiltonians, non-Hermitian Hamiltonians, negative probabilities, etc. None
of these operations really changes the theory: they just change the
reference frame, so to speak, within the same theory. Even the Lee-Wick idea
of making \textquotedblleft abnormal particles\textquotedblright\ decay \cite%
{leewick} cannot solve the problem\footnote{%
For Lee-Wick ghosts in quantum gravity, see \cite{tomboulis}.}, because a
theory with unstable ghosts is still a theory with ghosts. Necessarily, it
must be abandoned at some point, in favor of a different theory, and the
switch from one to the other must be a radical operation that cuts out the
sick portion, like a guillotine: this is the projection we are talking about.

The main application of the idea of purely virtual particle is the
formulation of a theory of quantum gravity \cite{LWgrav}, which provides
testable predictions \cite{ABP} in inflationary cosmology \cite{CMBStage4}.
In phenomenology, purely virtual particles open new possibilities, by
evading many constraints that are typical of normal particles (see \cite%
{PivaMelis} and references therein). The diagrammatic calculations are not
more difficult than those based on physical particles. It is possible to
implement them in softwares like FeynCalc, FormCalc, LoopTools and Package-X 
\cite{calc}.

Purely virtual particles can also be used as mere mathematical tools, to
study uncommon aspects of common theories, as shown in \cite%
{AbsoPhys,AbsoPhysGrav} and here. In this paper, we are using them to deal
with the local symmetries at finite $\tau $ and on a compact $\Omega $, to
express the initial, final and boundary conditions in invariant ways.

\bigskip

In common textbooks, the diagrammatic formulation of quantum field theory
focuses on the $S$ matrix ($\tau =\infty $), while the $\tau <\infty $ case
is mostly treated formally, from the operatorial definition $U(t_{\text{f}%
},t_{\text{i}})=\mathrm{e}^{-iH\tau }$. Not only, but when a compact space
manifold is considered, it is typically the torus, which does not pose
particular difficulties. Beyond the textbook approaches, and besides our
previous paper \cite{MQ}, we point out the results of Nomoto and Fukuda, who
studied QED at finite $\tau $ in ref. \cite{japan}, still on the torus. Yet,
the challenges of non-Abelian Yang-Mills theories and quantum gravity at $%
\tau <\infty $ on an arbitrary (especially, compact) space manifold $\Omega $
require the general formalism developed here.

The results of this paper and \cite{MQ} make us less dependent on the
paradigms that have dominated the scene in quantum field theory since its
birth. For example, we can study unitarity without being tied to the $S$
matrix. This is important in quantum gravity, where it makes no sense to
talk about the unitarity of the $S$ matrix (if the cosmological constant $%
\Lambda _{C}$ is nonvanishing), since proper definitions of asymptotic
states and $S$ matrix amplitudes are unavailable at $\Lambda _{C}\neq 0$.
Yet, unitarity is an essential requirement for a theory to be physically
acceptable. Other concerns revolve around the definition of energy and the
treatment of the Hamiltonian. Here we bypass these problems. We show that
the evolution operator $U(t_{\text{f}},t_{\text{i}})$ of quantum gravity
with purely virtual particles, defined by the functional integral, is
diagrammatically unitary for arbitrary $\tau <\infty $. This means that the
problems of the $S$ matrix at $\Lambda _{C}\neq 0$ are not inherent to the
issue of unitarity \textit{per se}.

The paper is organized as follows. In section \ref{toy} we consider a simple
warm-up toy model to illustrate some of the issues we need to face when we
want to find the right eigenfunctions for the expansions of the gauge
fields. In section \ref{cohegauge} we work out the general formalism for
coherent states in gauge theories. In section \ref{gaugetheories} we
rearrange the Lagrangian in Yang-Mills theories to make it ready for the
restriction to finite $\tau $ and compact $\Omega $. In section \ref%
{cohequad} we introduce coherent states in Yang-Mills theories at the
quadratic level. In section \ref{inter} we include the interactions. In
sections \ref{semiinfinitecylinder} and \ref{finitecylinder} we illustrate
the formalism in two relatively simple cases: Yang-Mills theory on the
semi-infinite cylinder, and on the finite cylinder. In section \ref{gravity}
we formulate Einstein gravity at finite $\tau $ and compact $\Omega $. In
section \ref{PVQG} we extend the formulation to quantum gravity with purely
virtual particles, and discuss the problems that occur in the limit $\tau
\rightarrow \infty $, $\Omega \rightarrow \mathbb{R}^{D-1}$, in the presence
of a cosmological constant. Section \ref{conclusions} contains the
conclusions.

\section{A warm-up toy model}

\label{toy}\setcounter{equation}{0}

The first difficulty we meet when we want to formulate gauge theories and
gravity on a compact manifold $\Omega $, is that we do not know the
eigenfunctions we should use for the expansions of the fields. In a generic
setting, the eigenfunctions of the Laplacian are not the right ones. In this
section we study a toy model that illustrates the main issue, as well as its
solution.

Specifically, we consider the simple quadratic Lagrangian%
\begin{equation}
L=\frac{\dot{\phi}^{2}}{2}+\alpha \dot{\phi}\phi ^{\prime }-\frac{\nu ^{2}}{2%
}\phi ^{\prime 2}  \label{lasta}
\end{equation}%
for a scalar field $\phi $ on a segment $\Omega =[0,\ell ]$ in a finite
interval of time $(t_{\text{i}},t_{\text{f}})$, $t_{\text{f}}=t_{\text{i}%
}+\tau $, with Dirichlet boundary conditions $\phi =0$ on $\partial \Omega $%
. The dot denotes the time derivative, while the prime denotes the space
derivative.

What is not clear is how to deal with the term $\dot{\phi}\phi ^{\prime }$.
We could eliminate it by means of a redefinition of space and time, but this
would complicate the investigation in another way, by mixing the boundary
conditions with the initial and final conditions. Moreover, we can apply the
redefinition only once (i.e., for a single field), which makes it useless in
the presence of more fields with kinetic Lagrangians of the same type. It is
necessary to work out a general approach that can be easily exported to the
cases treated in the next sections.

We begin by working out the momentum $\pi _{\phi }$ and the Hamiltonian $H$,
which are%
\begin{equation*}
\qquad \pi _{\phi }=\dot{\phi}+\alpha \phi ^{\prime },\qquad H(\pi _{\phi
},\phi )=\frac{1}{2}(\pi _{\phi }-\alpha \phi ^{\prime })^{2}+\frac{\nu ^{2}%
}{2}\phi ^{\prime 2}.
\end{equation*}%
Note that $H$ is positive definite for every real $\alpha $. Then we extend
the Lagrangian to%
\begin{equation}
L^{\prime }(\phi ,\dot{\phi},\pi _{\phi },\dot{\pi}_{\phi })=\frac{1}{2}(\pi
_{\phi }\dot{\phi}-\dot{\pi}_{\phi }\phi )-H(\pi _{\phi },\phi ),  \label{Lp}
\end{equation}%
which is convenient because it contains both $\phi $ and $\pi _{\phi }$ as
independent variables.

The equations of motion must be solved with the Dirichlet boundary
conditions $\phi =0$ on $\partial \Omega $. There is no boundary condition
on $\pi _{\phi }$, because, as we are going to see, the coherent states are
not built with $\phi $ and $\pi _{\phi }$, but with $\phi $ and $\dot{\phi}$%
. Note that $\phi =0$ on $\partial \Omega $ implies $\dot{\phi}=0$ on $%
\partial \Omega $. This way, the coherent states automatically vanish on $%
\partial \Omega $ as well. For these reasons, it is convenient to introduce
the shifted momenta%
\begin{equation}
\bar{\pi}_{\phi }=\pi _{\phi }-\alpha \phi ^{\prime },  \label{cano}
\end{equation}%
and add $\left. \bar{\pi}_{\phi }\right\vert _{\partial \Omega }=0$ to the
boundary conditions.

The integrated Lagrangian (\ref{Lp}) can be written as%
\begin{equation}
\mathcal{L}^{\prime }=\int_{0}^{\ell }L^{\prime }\mathrm{d}x=\frac{1}{2}%
\int_{0}^{\ell }\left( 
\begin{tabular}{ll}
$\bar{\pi}_{\phi }$ & $\phi $%
\end{tabular}%
\right) \left( 
\begin{tabular}{cc}
$-1$ & $\partial _{t}$ \\ 
$-\partial _{t}$ & $\nu ^{2}\partial _{x}^{2}-2\alpha \partial _{t}\partial
_{x}$%
\end{tabular}%
\right) \left( 
\begin{tabular}{l}
$\bar{\pi}_{\phi }$ \\ 
$\phi $%
\end{tabular}%
\right) \mathrm{d}x.  \label{inte}
\end{equation}%
The boundary conditions allow us to freely integrate by parts.

The field equations can be read from (\ref{inte}). The eigenfunctions with
energy $\omega $ ($\partial _{t}=-i\omega $) are%
\begin{equation}
\phi _{n}(x)=i\sqrt{\frac{2\nu ^{2}}{\ell (\nu ^{2}+\alpha ^{2})}}\exp
\left( -\frac{i\alpha \omega _{n}x}{\nu ^{2}}\right) \sin \left( \frac{n\pi x%
}{\ell }\right) ,\quad \bar{\pi}_{\phi _{n}}{}=-i\omega _{n}\phi _{n},\quad
\omega _{n}=\frac{n\pi \nu ^{2}}{\ell \sqrt{\nu ^{2}+\alpha ^{2}}},
\label{eigen}
\end{equation}%
having normalized them as explained below. We have $\phi _{n}^{\ast
}(x)=\phi _{-n}(x)$, $\omega _{-n}=-\omega _{n}$.

The expansions of the fields in terms of these eigenfunctions read%
\begin{equation}
\phi (t,x)=\sum_{n\neq 0}a_{n}(t)\phi _{n}(x),\qquad \bar{\pi}_{\phi
}(t,x)=-i\sum_{n\neq 0}a_{n}(t)\omega _{n}\phi _{n}(x),\qquad
a_{-n}(t)=a_{n}^{\ast }(t).  \label{expan}
\end{equation}%
The functional integral is the integral on the variables $a_{n}$ (or,
equivalently, the coherent states, see below). It is important to stress
that the expansions (\ref{expan}) \textit{define} the space of functions on
which the functional integral is calculated. In this spirit, we do not need
to prove, or require, that the expansions converge.

The orthogonality\ relations obeyed by the eigenfunctions can be worked out
as follows. From (\ref{inte}), we find%
\begin{equation*}
\left( 
\begin{tabular}{cc}
$-1$ & $-i\omega _{n}$ \\ 
$i\omega _{n}$ & $\nu ^{2}\partial _{x}^{2}+2i\alpha \omega _{n}\partial
_{x} $%
\end{tabular}%
\right) \left( 
\begin{tabular}{l}
$\bar{\pi}_{\phi _{n}}$ \\ 
$\phi _{n}$%
\end{tabular}%
\right) =0.
\end{equation*}%
Multiplying by the row $\left( 
\begin{tabular}{ll}
$\bar{\pi}_{\phi _{m}}$ & $\phi _{m}$%
\end{tabular}%
\right) $ and integrating on $\Omega $, we obtain%
\begin{equation}
0=\int_{0}^{\ell }\left( 
\begin{tabular}{ll}
$\bar{\pi}_{\phi _{m}}$ & $\phi _{m}$%
\end{tabular}%
\right) \left( 
\begin{tabular}{cc}
$-1$ & $-i\omega _{n}$ \\ 
$i\omega _{n}$ & $\nu ^{2}\partial _{x}^{2}+2i\alpha \omega _{n}\partial
_{x} $%
\end{tabular}%
\right) \left( 
\begin{tabular}{l}
$\bar{\pi}_{\phi _{n}}$ \\ 
$\phi _{n}$%
\end{tabular}%
\right) \mathrm{d}x.  \label{proda}
\end{equation}%
Transposing this expression, exchanging $n$ with $m$,\ integrating by parts
where necessary, and subtracting the result to (\ref{proda}), we find%
\begin{equation*}
0=(\omega _{n}+\omega _{m})\int_{0}^{\ell }\left( 
\begin{tabular}{ll}
$\bar{\pi}_{\phi _{m}}$ & $\phi _{m}$%
\end{tabular}%
\right) \left( 
\begin{tabular}{cc}
$0$ & $-1$ \\ 
$1$ & $2\alpha \partial _{x}$%
\end{tabular}%
\right) \left( 
\begin{tabular}{l}
$\bar{\pi}_{\phi _{n}}$ \\ 
$\phi _{n}$%
\end{tabular}%
\right) \mathrm{d}x.
\end{equation*}%
Dividing by $\omega _{n}+\omega _{m}$, we obtain the orthogonality relations
for $m\neq -n$. Choosing the normalization as in (\ref{eigen}), the
orthonormality relations read%
\begin{equation}
\int_{0}^{\ell }\left( 
\begin{tabular}{ll}
$\bar{\pi}_{\phi _{-m}}$ & $\phi _{-m}$%
\end{tabular}%
\right) \left( 
\begin{tabular}{cc}
$0$ & $1$ \\ 
$-1$ & $-2\alpha \partial _{x}$%
\end{tabular}%
\right) \left( 
\begin{tabular}{l}
$\bar{\pi}_{\phi _{n}}$ \\ 
$\phi _{n}$%
\end{tabular}%
\right) \mathrm{d}x=2i\omega _{n}\delta _{mn}.  \label{ecubo}
\end{equation}

Now we work out the expansion of the integrated Lagrangian (\ref{inte}).
Consider the right-hand side of the identity (\ref{proda}). Multiplying it
by $a_{m}a_{n}/2$, summing on $m$ and $n$, and adding the result to (\ref%
{inte}), we get%
\begin{equation*}
\mathcal{L}^{\prime }=\frac{1}{2}\sum_{n\neq 0,m\neq 0}a_{m}(\dot{a}%
_{n}+i\omega _{n}a_{n})\int_{0}^{\ell }\left( 
\begin{tabular}{ll}
$\bar{\pi}_{\phi _{m}}$ & $\phi _{m}$%
\end{tabular}%
\right) \left( 
\begin{tabular}{cc}
$0$ & $1$ \\ 
$-1$ & $-2\alpha \partial _{x}$%
\end{tabular}%
\right) \left( 
\begin{tabular}{l}
$\bar{\pi}_{\phi _{n}}$ \\ 
$\phi _{n}$%
\end{tabular}%
\right) \mathrm{d}x.
\end{equation*}%
Formula (\ref{ecubo}) ensures that all the terms with $m\neq -n$ drop out,
and we remain with%
\begin{equation*}
\mathcal{L}^{\prime }=\sum_{n>0}i\omega _{n}(a_{n}^{\ast }\dot{a}_{n}-\dot{a}%
_{n}^{\ast }a_{n})-2\sum_{n>0}\omega _{n}^{2}a_{n}^{\ast }a_{n},
\end{equation*}%
having halved the sum by using $a_{-n}=a_{n}^{\ast }$.

At this point, we define the coherent states $z_{n}=a_{n}$ and $\bar{z}%
_{n}=a_{n}^{\ast }$, and proceed as usual (see \cite{MQ} for a derivation in
the notation we are using here). Once we include the right endpoint
corrections, to have the correct variational problem, the complete action is%
\begin{equation}
S=-i\sum_{n>0}\omega _{n}(\bar{z}_{n\text{f}}z_{n}(t_{\text{f}})+\bar{z}%
_{n}(t_{\text{i}})z_{n\text{i}})+\sum_{n>0}\int_{t_{\text{i}}}^{t_{\text{f}}}%
\mathrm{d}t\left[ i\omega _{n}(\bar{z}_{n}\dot{z}_{n}-\dot{\bar{z}}%
_{n}z_{n})-2\omega _{n}^{2}\bar{z}_{n}z_{n}\right] ,  \label{Sz}
\end{equation}%
where $z_{n\text{i}}=z_{n}(t_{\text{i}})$, $\bar{z}_{n\text{f}}=\bar{z}%
_{n}(t_{\text{f}})$ are the initial and final conditions.

\section{Coherent states in gauge theories and gravity}

\label{cohegauge}\setcounter{equation}{0}

A nontrivial issue is to introduce coherent states in gauge theories and
gravity, and set invariant initial, final and boundary conditions. The goal
is to work in a general setting, which means without shortcuts (like
choosing particular gauge-fixings), because we want to have gauge
independence under control, and be able to make computations with arbitrary
gauge-fixing parameters, as we normally do at $\tau =\infty $, in $\Omega =%
\mathbb{R}^{D-1}$.

The properties we lay out in this section are useful for both gauge theories
and gravity, because they do not rely of the particular form of the local
symmetry. This is possible because, by means of the formalisms of refs. \cite%
{AbsoPhys,AbsoPhysGrav}, which we review in the next sections, we can
rephrase the local symmetries in a universal form, which amounts to
arbitrary shifts $\delta _{\Lambda }\varphi =\Lambda $ of certain (purely
virtual) extra fields $\varphi $. This is precisely the trick we need to
specify invariant conditions on the fields.

We start from a Lagrangian $L(\phi ,\dot{\phi})$ that depends on a certain
number of fields $\phi ^{I}$ and their first derivatives. We assume that it
can be decomposed as%
\begin{equation}
L(\phi ,\dot{\phi})=L_{\text{free}}(\phi ,\dot{\phi})+L_{\text{int}}(\phi ,%
\dot{\phi}),  \label{Lfreeint}
\end{equation}%
where $L_{\text{free}}$ is quadratic, and $L_{\text{int}}$ is the part to be
treated perturbatively (which may also include certain linear and quadratic
terms), to which we refer as \textquotedblleft interaction
Lagrangian\textquotedblright . For the moment, we assume that the boundary
conditions on the fields $\phi ^{I}$ are $\left. \phi ^{I}\right\vert
_{\partial \Omega }=0$. Nontrivial boundary conditions are studied at the
end of this section.

We assume that no Lagrangian term contains more than two derivatives.
Higher-derivative theories must be first turned into two-derivative theories
(by introducing extra fields, for example). Moreover, at finite $\tau $, on
a compact space manifold $\Omega $, we assume that terms like $\phi
_{1}\cdots \phi _{n-1}\partial \partial \phi _{n}$ have been eliminated in
favor of terms like $\phi _{1}\cdots \phi _{n-2}\partial \phi _{n-1}\partial
\phi _{n}$, by adding total derivatives. In the next sections we show how to
do these and other operations while preserving gauge invariance and general
covariance.

Next, we assume that $L$ it is \textquotedblleft orthodoxically
symmetric\textquotedblright\ with respect to certain infinitesimal
transformations $\delta _{\Lambda }\phi ^{I}$. By this we mean that

$i$) the functions $\delta _{\Lambda }\phi ^{I}$ depend only on the fields $%
\phi ^{I}$, but not on their derivatives,

$ii$) the Lagrangian satisfies%
\begin{equation}
0=\delta _{\Lambda }\phi ^{I}\frac{\partial L}{\partial \phi ^{I}}+\delta
_{\Lambda }\dot{\phi}^{I}\frac{\partial L}{\partial \dot{\phi}^{I}},
\label{gaugeida}
\end{equation}%
where%
\begin{equation}
\delta _{\Lambda }\dot{\phi}^{I}=\partial _{t}(\delta _{\Lambda }\phi ^{I})=%
\frac{\partial (\delta _{\Lambda }\phi ^{I})}{\partial \phi ^{J}}\dot{\phi}%
^{J}.  \label{deltafip}
\end{equation}

What is important, in point $ii$), is that not only the action is symmetric,
but also the Lagrangian is, i.e., the right-hand side of (\ref{gaugeida}) is
exactly zero, not just a total derivative.

Next, we introduce the momenta and the Hamiltonian as usual\footnote{%
In order to keep the notation simple, we adopt the following conventions for
fields with fermionic statistics. Once their kinetic terms are diagonalized,
we have pairs $\bar{\psi}$, $\psi $. The quadratic terms we write must be
understood as follows: $\bar{\psi}$ is placed to the left, and $\psi $ is
placed to the right; $\pi _{\psi }$ is defined as the left derivative with
respect to $\dot{\bar{\psi}}$, and is placed to the right; $\pi _{\bar{\psi}%
} $ is defined as the right derivative with respect to $\dot{\psi}$, and is
placed to the left.}:%
\begin{equation*}
\pi _{\phi }^{I}(\phi ,\dot{\phi})=\frac{\partial L}{\partial \dot{\phi}^{I}}%
\qquad \Rightarrow \qquad \dot{\phi}^{I}=\dot{\phi}^{I}(\pi _{\phi },\phi
)\equiv \dot{\tilde{\phi}}^{I},\qquad H(\pi _{\phi },\phi )=\pi _{\phi }^{I}%
\dot{\tilde{\phi}}^{I}-L(\phi ,\dot{\tilde{\phi}}).
\end{equation*}%
We can work out the symmetry transformations of the momenta $\pi _{\phi
}^{I} $ by means of the identities (\ref{gaugeida}) and (\ref{deltafip}). We
find%
\begin{equation}
\delta _{\Lambda }\pi _{\phi }^{I}=-\frac{\partial (\delta _{\Lambda }\dot{%
\phi}^{J})}{\partial \dot{\phi}^{I}}\pi _{\phi }^{J}=-\frac{\partial (\delta
_{\Lambda }\phi ^{J})}{\partial \phi ^{I}}\pi _{\phi }^{J}.  \label{deltap}
\end{equation}%
Since $\delta _{\Lambda }\dot{\phi}^{J}$ is linear in $\dot{\phi}^{I}$, $%
\delta _{\Lambda }\pi _{\phi }^{I}$ depends only on $\phi $ and $\pi _{\phi
} $, but not on $\dot{\phi}$.

We want to prove that the equivalent, extended Lagrangian%
\begin{equation*}
L^{\prime \prime }(\phi ,\dot{\phi},\pi _{\phi })=\pi _{\phi }^{I}\dot{\phi}%
^{I}-H(\pi _{\phi },\phi )=\pi _{\phi }^{I}(\dot{\phi}^{I}-\dot{\tilde{\phi}}%
^{I})+L(\phi ,\dot{\tilde{\phi}})
\end{equation*}%
is orthodoxically symmetric, the transformations being $\delta _{\Lambda
}\phi ^{I}$ and (\ref{deltap}).

Since the transformations $\delta _{\Lambda }\phi ^{I}$ and $\delta
_{\Lambda }\pi _{\phi }^{I}$ do not depend on the derivatives of the fields,
point $i$) is satisfied. It remains to prove the equation%
\begin{equation}
0=\delta _{\Lambda }\phi ^{I}\frac{\partial L^{\prime \prime }}{\partial
\phi ^{I}}+\delta _{\Lambda }\dot{\phi}^{I}\frac{\partial L^{\prime \prime }%
}{\partial \dot{\phi}^{I}}+\delta _{\Lambda }\pi _{\phi }^{I}\frac{\partial
L^{\prime \prime }}{\partial \pi _{\phi }^{I}}.  \label{dLp}
\end{equation}%
For this purpose, note that formula (\ref{gaugeida}) with the replacement $%
\dot{\phi}^{I}\rightarrow \dot{\tilde{\phi}}^{I}$ gives%
\begin{equation*}
0=\delta _{\Lambda }\phi ^{I}\left. \frac{\partial L(\phi ,\dot{\tilde{\phi}}%
^{I})}{\partial \phi ^{I}}\right\vert _{\dot{\tilde{\phi}}}+\pi _{\phi }^{I}%
\frac{\partial (\delta _{\Lambda }\phi ^{I})}{\partial \phi ^{J}}\dot{\tilde{%
\phi}}^{J},
\end{equation*}%
using (\ref{deltafip}). Then it is easy to check that the right-hand side of
the identity (\ref{dLp}) is equal to 
\begin{equation*}
(\dot{\phi}^{I}-\dot{\tilde{\phi}}^{I})\left( \delta _{\Lambda }\pi _{\phi
}^{I}+\frac{\partial (\delta _{\Lambda }\phi ^{J})}{\partial \phi ^{I}}\pi
_{\phi }^{J}\right) ,
\end{equation*}%
which vanishes by (\ref{deltap}).

We need to make a further step, because the extended Lagrangian we must
start from, in the coherent-state approach, is not $L^{\prime \prime }$, but 
\begin{equation}
L^{\prime }(\phi ,\dot{\phi},\pi _{\phi },\dot{\pi}_{\phi })=\frac{1}{2}%
\left( \pi _{\phi }^{I}\dot{\phi}^{I}-\dot{\pi}_{\phi }^{I}\phi ^{I}\right)
-H(\pi _{\phi },\phi )=L^{\prime \prime }-\frac{1}{2}\frac{\mathrm{d}}{%
\mathrm{d}t}\left( \pi _{\phi }^{I}\phi ^{I}\right) .  \label{Lpcohe}
\end{equation}%
We will also need to add certain endpoint corrections to the action, in
order to have the right variational problem. This part can be ignored for
the moment, because it will be easy to deal with it at the very end.

It is not obvious that the total derivative $L^{\prime }-L^{\prime \prime }$
is invariant under the transformation $\delta _{\Lambda }$. Actually, in
general it is not, since (\ref{deltap}) gives%
\begin{equation}
\delta _{\Lambda }(\pi _{\phi }^{I}\phi ^{I})=\pi _{\phi }^{I}\left( \delta
_{\Lambda }\phi ^{I}-\frac{\partial (\delta _{\Lambda }\phi ^{I})}{\partial
\phi ^{J}}\phi ^{J}\right) ,  \label{quad}
\end{equation}%
which vanishes only if the transformations are linear:%
\begin{equation}
\delta _{\Lambda }\phi ^{I}=\frac{\partial (\delta _{\Lambda }\phi ^{I})}{%
\partial \phi ^{J}}\phi ^{J}.  \label{lincond}
\end{equation}%
Summarizing, if the symmetry is linear, the Lagrangian (\ref{Lpcohe}) is
orthodoxically invariant.

It may seem that the requirement of having linear symmetry transformations
is very restrictive. Actually, it is not, if we take advantage of the
formalism developed in refs. \cite{AbsoPhys,AbsoPhysGrav}. Indeed, it is
always possible to convert Abelian and non-Abelian gauge symmetries, as well
as general covariance, into a universal linear form, by introducing purely
virtual fields that do not change the $S$ matrix amplitudes.

It is easy to check that the momenta $\pi _{\phi }^{I}$ are not guaranteed
to vanish on $\partial \Omega $. The structure of the Lagrangian ensures
that $\pi _{\phi }^{I}(\phi ,\dot{\phi})$ has the form%
\begin{equation*}
\pi _{\phi }^{I}(\phi ,\dot{\phi})=\mathcal{A}^{IJ}(\phi )\dot{\phi}^{J}+%
\mathcal{B}^{IJi}(\phi )\partial _{i}\phi ^{J}+\mathcal{C}^{I}(\phi ),
\end{equation*}%
for certain functions $\mathcal{A}^{IJ}(\phi )$, $\mathcal{B}^{IJi}(\phi )$
and $\mathcal{C}^{I}(\phi )$. Thus, $\left. \phi ^{I}\right\vert _{\partial
\Omega }=0$ implies%
\begin{equation}
\left. \pi _{\phi }^{I}(\phi ,\dot{\phi})\right\vert _{\partial \Omega }=%
\mathcal{B}^{IJi}(0)\left. \partial _{i}\phi ^{J}\right\vert _{\partial
\Omega }+\mathcal{C}^{I}(0).  \label{pide}
\end{equation}%
We can assume $\mathcal{C}^{I}(0)=0$. First, note that a nonvanishing $%
\mathcal{C}^{I}(0)$ means that the Lagrangian includes a term $\mathcal{C}%
^{I}(0)\dot{\phi}^{I}$. This is not going to happen in the cases of
Yang-Mills theories and gravity. Besides, a term like $\mathcal{C}^{I}(0)%
\dot{\phi}^{I}$ can be removed at no cost. Since we are assuming that the
symmetry transformations are linear and do not involve derivatives, $%
\mathcal{C}^{I}(0)\dot{\phi}^{I}$ must be gauge invariant by itself.
Besides, it is a total derivative. Thus, we can always switch to an
alternative Lagrangian with the same properties, but no such term. Instead,
the matrix $\mathcal{B}^{IJi}(0)$ is in general nontrivial and cannot be
removed, so the right-hand side of (\ref{pide}) may be nonzero.

As in (\ref{cano}), it is useful to define new \textquotedblleft
momenta\textquotedblright 
\begin{equation}
\bar{\pi}_{\phi }^{I}=\pi _{\phi }^{I}-\mathcal{B}^{IJi}(0)\partial _{i}\phi
^{J},  \label{pipi}
\end{equation}%
because then it makes sense to add the boundary conditions%
\begin{equation}
\left. \left. \phi ^{I}\right\vert _{\partial \Omega }=\bar{\pi}_{\phi
}^{I}\right\vert _{\partial \Omega }=0.  \label{bonda}
\end{equation}%
As we show below, these conditions turn straightforwardly into the right
boundary conditions for the coherent states.

The gauge transformations of $\bar{\pi}_{\phi }^{I}$ follow from those of $%
\pi _{\phi }^{I}$ and $\phi ^{I}$. This is enough, for the moment, but in
subsection \ref{gtcohe} we prove $\pi _{\phi }^{I}$ and $\bar{\pi}_{\phi
}^{I}$ transform in exactly the same way.

By assumption (\ref{Lfreeint}) and the absence of higher derivatives, the
general form of the Lagrangian $L^{\prime }$ is%
\begin{equation}
L^{\prime }=L_{\text{free}}^{\prime }(\phi ,\dot{\phi},\bar{\pi}_{\phi },%
\dot{\bar{\pi}}_{\phi })+L_{\text{int}}^{\prime }(\bar{\pi}_{\phi },\phi ),
\label{linva}
\end{equation}%
where $L_{\text{free}}^{\prime }$ is quadratic, and the interaction part $L_{%
\text{int}}^{\prime }$ is independent of the time derivatives. Note that the
redefinitions (\ref{pipi}) do not generate time derivatives in the
interaction sector. The quadratic Lagrangian, integrated on $\Omega $, has
the form 
\begin{equation}
\qquad \mathcal{L}_{\text{free}}^{\prime }\equiv \int_{\Omega }L_{\text{free}%
}^{\prime }\mathrm{d}^{D-1}\bm{x}\equiv \frac{1}{2}\int_{\Omega }\left( 
\begin{tabular}{ll}
$\bar{\pi}_{\phi }$ & $\phi $%
\end{tabular}%
\right) \left( 
\begin{tabular}{cc}
$M$ & $K+\partial _{t}$ \\ 
$\tilde{K}-\partial _{t}$ & $N$%
\end{tabular}%
\right) \left( 
\begin{tabular}{c}
$\bar{\pi}_{\phi }$ \\ 
$\phi $%
\end{tabular}%
\right) \mathrm{d}^{D-1}\bm{x},  \label{lip}
\end{equation}%
where $M$ is a constant, symmetric matrix, while $K=K_{1}^{i}\partial
_{i}+K_{2}$, $\tilde{K}=-K_{1}^{iT}\partial _{i}+K_{2}^{T}$ ($K_{1}^{i}$ and 
$K_{2}$ being matrices, $T$ denoting the transpose), $N=N_{1}^{ij}\partial
_{i}\partial _{j}+N_{2}^{i}\partial _{i}\partial _{t}+N_{3}^{i}\partial
_{i}+N_{4}$, with $N_{1}^{ij}$, $N_{2}^{i}$, $N_{4}$ symmetric matrices and $%
N_{3}^{i}$ antisymmetric. Observe that, by (\ref{bonda}), we can freely
integrate the space derivatives by parts.

\subsection{Frequencies and eigenfunctions}

\label{freque}

The eigenfunctions $\bar{\pi}_{n}^{I}(\bm{x)}$, $\phi _{n}^{I}(\bm{x)}$ are
the solutions of the problem 
\begin{equation}
\left( 
\begin{tabular}{cc}
$M$ & $K_{1}^{i}\partial _{i}+K_{2}-i\omega _{n}$ \\ 
$i\omega _{n}-K_{1}^{iT}\partial _{i}+K_{2}^{T}$ & $N_{1}^{ij}\partial
_{i}\partial _{j}-i\omega _{n}N_{2}^{i}\partial _{i}+N_{3}^{i}\partial
_{i}+N_{4}$%
\end{tabular}%
\right) \left( 
\begin{tabular}{c}
$\bar{\pi}_{n}$ \\ 
$\phi _{n}$%
\end{tabular}%
\right) =0,  \label{eigeneq}
\end{equation}%
with the boundary conditions $\left. \bar{\pi}_{n}^{I}\right\vert _{\partial
\Omega }=\left. \phi _{n}^{I}\right\vert _{\partial \Omega }=0$, where $n$
is some label.

We assume that the frequencies are real, because they are so in the
applications we have in mind. A quick proof is as follows. The frequencies
are real for $\tau =\infty $, $\Omega =\mathbb{R}^{D-1}$, in both Yang-Mills
theory and gravity. Let us denote them by $\omega _{\infty }$. We can work
out the frequencies $\omega _{n}$ and the eigenfunctions at finite $\tau $,
compact $\Omega $, by considering linear combinations of the $\tau =\infty $%
, $\Omega =\mathbb{R}^{D-1}$ eigenfunctions with identical frequencies $%
\omega _{\infty }$, and fixing the coefficients by means of the boundary
conditions. Eventually, the frequencies become discrete, to have solutions,
but remain real.

In case of need, it is not difficult to generalize the formulas of this
paper to complex frequencies. We just remark that they must appear in
complex conjugate pairs, since the Lagrangian is assumed to be Hermitian.

Taking the complex conjugate of (\ref{eigeneq}), we find that $\bar{\pi}%
_{n}^{I\ast }(\bm{x})$ and $\phi _{n}^{I\ast }(\bm{x})$ are also
eigenfunctions, and their frequency is $-\omega _{n}$. We use $n^{\ast }$ to
label them, and write%
\begin{equation}
\bar{\pi}_{n^{\ast }}^{I}(\bm{x})=\bar{\pi}_{n}^{I\ast }(\bm{x}),\qquad \phi
_{n^{\ast }}^{I}(\bm{x})=\phi _{n}^{I\ast }(\bm{x}),\qquad \omega _{n^{\ast
}}=-\omega _{n}.  \label{pif}
\end{equation}%
If $\mathcal{V}$ denotes the range of the label $n$, we write $\mathcal{V=U}%
\cup \mathcal{U}^{\ast }$, by splitting each pair $n,n^{\ast }$ between $%
\mathcal{U}$ and $\mathcal{U}^{\ast }$.

The orthogonality relations can be worked out as in section \ref{toy}: $i$)
we multiply (\ref{eigeneq}) by $\left( \bar{\pi}_{m},\phi _{m}\right) $ and
integrate the product on $\Omega $; $ii$) we transpose the result of $i$),
exchange $n$ with $m$, and integrate by parts where necessary; finally, $iii$%
) we subtract the results of $i$) and $ii$).

Normalizing the eigenfunctions appropriately, we have the orthonormality
relations%
\begin{equation}
\int_{\Omega }\mathrm{d}^{D-1}\bm{x}\left( 
\begin{tabular}{ll}
$\bar{\pi}_{m}$ & $\phi _{m}$%
\end{tabular}%
\right) \left( 
\begin{tabular}{cc}
$0$ & $1$ \\ 
$-1$ & $N_{2}^{i}\partial _{i}$%
\end{tabular}%
\right) \left( 
\begin{tabular}{c}
$\bar{\pi}_{n}$ \\ 
$\phi _{n}$%
\end{tabular}%
\right) =2i\mathbb{\tau }_{n}\omega _{n}\delta _{m,n^{\ast }},
\label{ortono}
\end{equation}%
where $\mathbb{\tau }_{n}=\pm 1=\mathbb{\tau }_{n^{\ast }}$. The value $%
\mathbb{\tau }_{n}=-1$ signals the presence of ghosts (fields with kinetic
terms multiplied by the wrong signs). Indeed, going through the toy model of
the previous section, it is easy to check that, if we change the overall
sign of the starting Lagrangian (\ref{lasta}), the right-hand side of (\ref%
{ecubo}) turns out to be equal to $-2i\omega _{n}\delta _{mn}$.

We then expand $\bar{\pi}_{n}$ and $\phi _{n}$ in the basis we have just
worked out:%
\begin{equation}
\left( 
\begin{tabular}{c}
$\bar{\pi}_{\phi }$ \\ 
$\phi $%
\end{tabular}%
\right) =\sum_{n\in \mathcal{V}}a_{n}\left( 
\begin{tabular}{c}
$\bar{\pi}_{n}$ \\ 
$\phi _{n}$%
\end{tabular}%
\right) ,  \label{expanda}
\end{equation}%
with $a_{n^{\ast }}=a_{n}^{\ast }$. By means of (\ref{ortono}), we can
invert the expansion and find the coefficients:%
\begin{equation}
-2i\mathbb{\tau }_{m}\omega _{m}a_{m}^{\ast }(t)=\int_{\Omega }\mathrm{d}%
^{D-1}\bm{x}\left( 
\begin{tabular}{ll}
$\bar{\pi}_{m}(\bm{x)}$ & $\phi _{m}(\bm{x)}$%
\end{tabular}%
\right) \left( 
\begin{tabular}{cc}
$0$ & $1$ \\ 
$-1$ & $N_{2}^{i}\partial _{i}$%
\end{tabular}%
\right) \left( 
\begin{tabular}{c}
$\bar{\pi}_{\phi }(t,\bm{x)}$ \\ 
$\phi (t,\bm{x)}$%
\end{tabular}%
\right) .  \label{coeffa}
\end{equation}

We insert (\ref{expanda}) into (\ref{lip}), and then subtract (\ref{eigeneq}%
), multiplied by $(\bar{\pi}_{m},\phi _{m})a_{m}/2$, summed on $m,n\in 
\mathcal{V}$ and integrated on $\Omega $. Then, we use (\ref{ortono}), and
mirror the sum on $\mathcal{U}^{\ast }$ into a sum on $\mathcal{U}$. The
result is the integrated free Lagrangian%
\begin{equation}
\mathcal{L}_{\text{free}}^{\prime }=\sum_{n\in \mathcal{U}}i\mathbb{\tau }%
_{n}\omega _{n}(a_{n}^{\ast }\dot{a}_{n}-\dot{a}_{n}^{\ast
}a_{n})-2\sum_{n\in \mathcal{U}}\mathbb{\tau }_{n}\omega _{n}^{2}a_{n}^{\ast
}a_{n}.  \label{Lip}
\end{equation}

If the fields $\phi ^{I}$ have, say, $r$ independent components, $I=1,\cdots
,r$, the solutions of the eigenvalue problem can be arranged into $r$
independent, complete sets of eigenfunctions, each of which can be assigned
to a specific component $\phi ^{I}$. We can split the set $\mathcal{U}$ into
a union $\cup _{I=1}^{r}\mathcal{U}^{I}$, where $\mathcal{U}^{I}$ refers to
the $I$-th complete set. For convenience, we relabel the indices $n$ so that
their range is the same for each $I$, to be denoted by $\mathcal{\hat{U}}$.

Let $\bar{\pi}_{n}^{IJ}$ and $\phi _{n}^{IJ}$ denote the $I$-th components
of the $n$-th eigenfunction of the $J$-th set. Let $z_{n}(t)=a_{n}(t)$
denote the column made by $z_{n}^{I}(t)=a_{n}^{I}(t)$, $I=1,\cdots ,r$. We
have, from (\ref{expanda}),%
\begin{equation}
\left( 
\begin{tabular}{c}
$\bar{\pi}_{\phi }$ \\ 
$\phi $%
\end{tabular}%
\right) =\sum_{n\in \mathcal{\hat{U}}}\left( 
\begin{tabular}{ll}
$\bar{\pi}_{n}^{\ast }$ & $\bar{\pi}_{n}$ \\ 
$\phi _{n}^{\ast }$ & $\phi _{n}$%
\end{tabular}%
\right) \left( 
\begin{tabular}{c}
$\bar{z}_{n}$ \\ 
$z_{n}$%
\end{tabular}%
\right) ,  \label{coheba}
\end{equation}%
where $\bar{\pi}_{n}$, $\phi _{n}$, denote the block matrices made by $\bar{%
\pi}_{n}^{IJ}$ and $\phi _{n}^{IJ}$, while $\bar{\pi}_{n}^{\ast }$ and $\phi
_{n}^{\ast }$ are the conjugate matrices. The coefficients $\bar{z}_{n}^{I}$%
, $z_{n}^{I}$ of the expansion are the variables we call coherent
\textquotedblleft states\textquotedblright . The inverse formula reads, from
(\ref{coeffa}), 
\begin{equation}
-2i\mathbb{\tau }_{n}\omega _{n}\left( 
\begin{tabular}{l}
$\bar{z}_{n}$ \\ 
$z_{n}$%
\end{tabular}%
\right) =\int_{\Omega }\mathrm{d}^{D-1}\bm{x}\left( 
\begin{tabular}{cc}
$\bar{\pi}_{n}$ & $\phi _{n}$ \\ 
$-\bar{\pi}_{n}^{\ast }$ & $-\phi _{n}^{\ast }$%
\end{tabular}%
\right) \left( 
\begin{tabular}{cc}
$0$ & $1$ \\ 
$-1$ & $N_{2}^{i}\partial _{i}$%
\end{tabular}%
\right) \left( 
\begin{tabular}{c}
$\bar{\pi}_{\phi }$ \\ 
$\phi $%
\end{tabular}%
\right) .  \label{coeffaz}
\end{equation}

We can rearrange (\ref{Lip}) as%
\begin{equation}
\mathcal{L}_{\text{free}}^{\prime }=\sum_{I=1}^{r}\sum_{n\in \mathcal{\hat{U}%
}}\mathbb{\tau }_{n}^{I}\left[ i\omega _{n}^{I}(\bar{z}_{n}^{I}\dot{z}%
_{n}^{I}-\dot{\bar{z}}_{n}^{I}z_{n}^{I})-2\omega _{n}^{I\hspace{0.01in}2}%
\bar{z}_{n}^{I}z_{n}^{I}\right] .  \label{Lipfree}
\end{equation}%
Typically, the $\mathbb{\tau }_{n}^{I}$ factor we see here does not depend
on $n$, but just on $I$.

At this point, it is straightforward to add the interacting Lagrangian $L_{%
\text{int}}^{\prime }(\bar{\pi}_{\phi },\phi )$. We recall that $L_{\text{int%
}}^{\prime }$ does not contain time derivatives of $\bar{\pi}_{\phi }$ and $%
\phi $, by construction, although it can contain space derivatives.
Expanding the fields and the momenta in the basis (\ref{coheba}) of coherent
states, and integrating by parts when needed, we obtain an integrated
interacting Lagrangian%
\begin{equation*}
\mathcal{L}_{\text{int}}^{\prime }=\int_{\Omega }L_{\text{int}}^{\prime }%
\mathrm{d}^{D-1}\bm{x}
\end{equation*}%
that just depends $\bar{z}_{n}^{I}$, $z_{n}^{I}$ (no time derivatives).

Finally, the total action is%
\begin{equation}
S=-i\sum_{I=1}^{r}\sum_{n\in \mathcal{\hat{U}}}\mathbb{\tau }_{n}^{I}\omega
_{n}^{I}\left( \bar{z}_{n\text{f}}^{I}z_{n}^{I}(t_{\text{f}})+\bar{z}%
_{n}^{I}(t_{\text{i}})z_{n\text{i}}^{I}\right) +\int_{t_{\text{i}}}^{t_{%
\text{f}}}\mathrm{d}t\left( \mathcal{L}_{\text{free}}^{\prime }+\mathcal{L}_{%
\text{int}}^{\prime }\right) ,  \label{action}
\end{equation}%
where $z_{n\text{i}}^{I}=$ $z_{n}^{I}(t_{\text{i}})$ parametrize the initial
conditions in the coherent-state approach, while $\bar{z}_{n\text{f}}^{I}=%
\bar{z}_{n}^{I}(t_{\text{f}})$ parametrize the final conditions. The sums
appearing in (\ref{action}), which we call \textquotedblleft endpoint
corrections\textquotedblright , are there to have the correct variational
problem. This means that the variations $\delta \bar{z}_{n}^{I}$, $\delta
z_{n}^{I}$, subject to the initial and final conditions $\delta \bar{z}%
_{n}^{I}(t_{\text{f}})=\delta z_{n}^{I}(t_{\text{i}})=0$, must give the $%
\bar{z}_{n}^{I}$ and $z_{n}^{I}$ equations of motion, and no further
restrictions. Note that the time derivatives of $\bar{z}_{n}^{I}$ and $%
z_{n}^{I}$ appear only inside $\mathcal{L}_{\text{free}}^{\prime }$. This is
the reason why the partial integrations that take care of the terms
proportional to $\delta \dot{\bar{z}}_{n}^{I}$ and $\delta \dot{z}_{n}^{I}$
are compensated by endpoint corrections as simple as those of (\ref{action}).

\subsection{Gauge transformations of coherent states}

\label{gtcohe}

Now we study the gauge transformations of the coherent states, and the
conditions to have gauge invariant amplitudes. As usual, the parameters $%
\Lambda $ of the gauge transformations are written as $\Lambda =\theta C$,
where $\theta $ is a constant, anticommuting parameter and $C$ are the
Faddeev-Popov ghosts. The fields $\phi ^{I}$ include $C$ and the other
fields that are necessary to gauge-fix the theory, which are the antighosts $%
\bar{C}$ and certain Lagrange multipliers $B$ for the gauge-fixing (see
below).

Since we are assuming the linearity conditions (\ref{lincond}), we can write
the gauge transformations as $\delta \phi ^{I}=\theta \Sigma ^{IJ}\phi ^{J}$%
, for some constants $\Sigma ^{IJ}$. By means of linear field redefinitions,
we can always split the set of fields $\phi ^{I}$ into three subsets $\phi
^{I_{+}}$, $\phi ^{I_{-}}$ and $\phi ^{I_{0}}$, where: $i$) the fields $\phi
^{I_{+}}$ transform into other fields; $ii$) the fields $\phi ^{I_{-}}$
parametrize the transformations of other fields; and $iii$) the fields $\phi
^{I_{0}}$ are invariant and cannot be obtained from the transformations of
other fields: 
\begin{equation}
\delta \phi ^{I_{+}}=\theta \phi ^{I_{-}},\qquad \delta \phi
^{I_{-}}=0,\qquad \delta \phi ^{I_{0}}=0.  \label{smila}
\end{equation}%
The transformation law can be written as 
\begin{equation*}
\delta =\delta \phi ^{I_{+}}\frac{\delta _{l}}{\delta \phi ^{I_{+}}}=\theta
\Delta ,\qquad \Delta \equiv \phi ^{I_{-}}\frac{\delta _{l}}{\delta \phi
^{I_{+}}},
\end{equation*}%
where $\delta _{l}$ denotes the left functional derivative.

The operator $\Delta $ has a standard \textquotedblleft
descent\textquotedblright\ structure. A well-known theorem (see appendix \ref%
{DeltaProp} for a direct proof) says that the most general solution of the
problem $\delta X=0$, where $X$ is a local function, is 
\begin{equation}
X=X_{0}+\Delta Y,  \label{teo}
\end{equation}%
where $X_{0}$ is a $\phi ^{I_{\pm }}$ independent local function, and $Y$ is
a local function.

Consider the invariant quadratic terms that we can build with the fields $%
\phi ^{I_{\pm }}$. At some point, we may need to diagonalize them. It is
easy to see that we cannot build enough invariant terms, unless the
diagonalization organizes the field $\phi ^{I_{\pm }}$ in \textquotedblleft
pairs of pairs\textquotedblright . Consider a single pair $\phi ^{I_{\pm }}$%
, and observe that $\phi ^{I_{+}}$ and $\phi ^{I_{-}}$ have opposite
statistics. By (\ref{teo}), the quadratic terms in question must be
contained in $\Delta Y$. However, the expressions $\Delta (\phi ^{I_{+}}\phi
^{I_{+}})$, $\Delta (\phi ^{I_{+}}\phi ^{I_{-}})$ and $\Delta (\phi
^{I_{-}}\phi ^{I_{-}})$ generate just one independent quadratic term, while
we need two. This means that for each pair $\phi ^{I_{\pm }}$ there must be
another pair $\phi ^{I_{\pm }\prime }$, out of which the required invariants
can be built.

We can organize the fields $\phi ^{I_{\pm }}$, $\phi ^{I_{\pm }\prime }$
into doublets. Using a notation that is ready for the applications to
Yang-Mills theories and gravity (adapting the meaning of the index $a$), we
write the doublets as 
\begin{equation}
\phi _{+}^{a}=\left( 
\begin{tabular}{l}
$\phi ^{a}$ \\ 
$\bar{C}^{a}$%
\end{tabular}%
\right) ,\qquad \phi _{-}^{a}=\left( 
\begin{tabular}{l}
$B^{a}$ \\ 
$C^{a}$%
\end{tabular}%
\right) ,  \label{smilo}
\end{equation}%
where $\phi ^{a}$ and $B^{a}$ have bosonic statistics, while $\bar{C}^{a}$
and $C^{a}$ have fermionic statistics. In all the applications that we have
in mind, this is the structure we need.

We can write the transformation law as%
\begin{equation}
\delta \phi _{+}^{a}=\theta \sigma _{1}\phi _{-}^{a},\qquad \delta =\theta
\Delta ,\qquad \Delta =\phi _{-}^{aT}\sigma _{1}\frac{\delta _{l}}{\delta
\phi _{+}^{a}},  \label{smilad}
\end{equation}%
where $\sigma _{1}$ is the first Pauli matrix and the superscript
\textquotedblleft $T$\textquotedblright\ means \textquotedblleft
transpose\textquotedblright .

The $\phi _{+}^{a}$ expansions (\ref{coheba}),%
\begin{equation*}
\phi _{+}^{a}=\sum_{J=1}^{r}\sum_{n\in \mathcal{\hat{U}}}(\phi
_{+n}^{aJ}z_{n}^{J}+\phi _{+n}^{aJ\ast }\bar{z}_{n}^{J}),
\end{equation*}%
must turn into the $\phi _{-}^{a}$ expansions under (\ref{smilad}):%
\begin{equation*}
\delta \phi _{+}^{a}=\sum_{J=1}^{r}\sum_{n\in \mathcal{\hat{U}}}(\phi
_{+n}^{aJ}\delta z_{n}^{J}+\phi _{+n}^{aJ\ast }\delta \bar{z}%
_{n}^{J})=\theta \sigma _{1}\phi _{-}^{a}=\theta \sum_{J=1}^{r}\sum_{n\in 
\mathcal{\hat{U}}}(\sigma _{1}\phi _{-n}^{aJ}z_{n}^{J}+\sigma _{1}\phi
_{-n}^{aJ\ast }\bar{z}_{n}^{J}).
\end{equation*}%
Since the equations (\ref{eigeneq}) are invariant under the symmetry, the
eigenfunctions appearing in the $\phi _{-}^{a}$ expansions must match
eigenfunctions appearing in the $\phi _{+}^{a}$ expansions. That is to say,
we must have $\phi _{+n}^{aJ_{+}}=\sigma _{1}\phi _{-n}^{aJ_{-}}$ for some
pairings of indices $J_{+}$, $J_{-}$. Then, the transformations of the
coherent states read $\delta z_{n}^{J_{+}}=\theta z_{n}^{J_{-}}$, $\delta
z_{n}^{J_{-}}=0$. Moreover, the $z_{n}^{J}$ with such indices must also be
organized in doublets, for the reasons explained above. Finally, the $\phi
^{I_{0}}$ expansions identify the invariant coherent states $z_{n}^{J_{0}}$.
We illustrate these facts in section \ref{cohequad}, formulas (\ref{bcc})
and (\ref{cotra}).

Summarizing, we can split the set of coherent states $z_{n}^{I}$ into three
subsets $w_{n}^{\alpha }$, $u_{n}^{a}$ and $v_{n}^{a}$ (with indices $\alpha 
$, $a$ spanning appropriate ranges). Both $u_{n}^{a}$ and $v_{n}^{a}$ are
doublets, with bosonic first components and fermionic second components.
Their gauge transformations are 
\begin{equation}
\delta w_{n}^{\alpha }=\delta \bar{w}_{n}^{\alpha }=0,\qquad \delta
u_{n}^{a}=\theta \sigma _{1}v_{n}^{a},\qquad \delta \bar{u}_{n}^{a}=\theta
\sigma _{1}\bar{v}_{n}^{a},\qquad \delta v_{n}^{a}=\delta \bar{v}_{n}^{a}=0.
\label{gaucohe}
\end{equation}

We can obtain results that agree with the ones just found by repeating the
analysis for the \textquotedblleft momenta\textquotedblright\ $\bar{\pi}%
_{\phi }^{I}$. The redefinition (\ref{pipi}) is due to the presence of the
terms $\sim \dot{\phi}^{I}\partial _{i}\phi ^{J}$ in the Lagrangian. Since
the symmetry is orthodox and linear, the sum of these terms must be gauge
invariant by itself. Taking into account the conventions we adopted for the
fields $\bar{\psi}$, $\psi $ with fermionic statistics, we can write such a
sum as%
\begin{equation*}
\dot{\phi}^{I}\mathcal{\tilde{B}}^{IJi}(0)\partial _{i}\phi ^{J},
\end{equation*}%
where $\mathcal{\tilde{B}}^{IJi}=\mathcal{B}^{IJi}$ if the indices $I$, $J$
refer to bosonic fields, or $I$ refers to $\bar{\psi}$, while $\mathcal{%
\tilde{B}}^{IJi}=-\mathcal{B}^{IJi}$ if $I$ refers to $\psi $. This way, the
redefinitions match (\ref{pipi}) precisely. Writing the transformations $%
\delta _{\Lambda }\phi ^{I}=\theta \Sigma ^{IJ}\phi ^{J}$, as above, gauge
invariance gives the condition%
\begin{equation*}
\Sigma ^{KI}\mathcal{\tilde{B}}^{KJi}(0)+(-1)^{\epsilon _{I}}\mathcal{\tilde{%
B}}^{IKi}(0)\Sigma ^{KJ}=0,
\end{equation*}%
where $\epsilon _{I}$ is the statistics of $\phi ^{I}$. Analyzing all the
situations one by one, we can easily see that this condition is equivalent
to 
\begin{equation*}
\Sigma ^{KI}\mathcal{B}^{KJi}(0)+\mathcal{B}^{IKi}(0)\Sigma ^{KJ}=0,
\end{equation*}%
which also gives the implication%
\begin{equation}
\delta _{\Lambda }\pi _{\phi }^{I}=-\Sigma ^{JI}\pi _{\phi }^{J}\qquad
\Rightarrow \qquad \delta _{\Lambda }\bar{\pi}_{\phi }^{I}=-\Sigma ^{JI}\bar{%
\pi}_{\phi }^{J},  \label{deltapp}
\end{equation}%
from (\ref{deltap}). Thus, the old and new momenta $\pi _{\phi }^{I}$ and $%
\bar{\pi}_{\phi }^{I}$ transform the same way.

The $\bar{\pi}_{\phi }^{I}$ expansions and their transformations can be
studied as we did for the fields $\phi ^{I}$. Matching the eigenfunctions,
we find agreement with (\ref{gaucohe}). Alternatively, we can study the
expansions of $\phi ^{I}$ and $\bar{\pi}_{\phi }^{I}$ at the same time by
working directly on (\ref{coheba}).

\bigskip

Since the Lagrangian (\ref{Lpcohe}) in gauge invariant under the
transformations $\delta _{\Lambda }\phi ^{I}$ and (\ref{deltap}), and we are
assuming the linearity conditions (\ref{lincond}), the Lagrangian (\ref%
{linva}) is invariant under $\delta _{\Lambda }\phi ^{I}$ and (\ref{deltapp}%
). The integrated Lagrangian $\mathcal{L}_{\text{free}}^{\prime }+\mathcal{L}%
_{\text{int}}^{\prime }$ of formula (\ref{action}) is invariant under (\ref%
{gaucohe}), once it is written in the variables $w_{n}^{\alpha }$, $%
u_{n}^{a} $ and $v_{n}^{a}$ and their conjugates.

The action (\ref{action})\ is gauge invariant if the endpoint corrections
are invariant, which occurs if they do not contain $u_{n}^{a}$ and $\bar{u}%
_{n}^{a}$. In addition, we require that they do not contain gauge trivial
modes, which are $v_{n}^{a}$ and $\bar{v}_{n}^{a}$ (which can be obtained as
transformations of $u_{n}^{a}$ and $\bar{u}_{n}^{a}$).

Thus, the physical amplitudes are those that have 
\begin{equation}
\bar{u}_{n\text{f}}^{a}=\bar{v}_{n\text{f}}^{a}=u_{n\text{i}}^{a}=v_{n\text{i%
}}^{a}=0,  \label{endrestr}
\end{equation}%
in which case the endpoint corrections, which read%
\begin{equation*}
-i\sum_{\alpha }\sum_{n\in \mathcal{\hat{U}}}\mathbb{\tau }_{n}^{\alpha
}\omega _{n}^{\alpha }\left( \bar{w}_{n\text{f}}^{\alpha }w_{n}^{\alpha }(t_{%
\text{f}})+\bar{w}_{n}^{\alpha }(t_{\text{i}})w_{n\text{i}}^{\alpha }\right)
,
\end{equation*}%
are manifestly gauge invariant.

The restrictions (\ref{endrestr}) on the endpoint corrections are analogous
to the restrictions we commonly apply to the $S$ matrix amplitudes: we do
not consider scattering processes involving Faddeev-Popov ghosts, or the
temporal and longitudinal components of the gauge fields, among the incoming
and outgoing states. Yet, sometimes it may\ be useful to relax these
requirements, and consider diagrams with all sorts of external legs,
including the ones just mentioned, to study renormalization, for example, or
the gauge independence of the physical quantities, or the diagrammatic
versions of the unitarity equations.

We conclude this subsection by writing down the universal structure of the
kinetic terms of the coherent states, inside $\mathcal{L}_{\text{free}%
}^{\prime }$. The ones of the gauge invariant sector are clearly%
\begin{equation}
\sum_{\alpha }\sum_{n\in \mathcal{\hat{U}}}\mathbb{\tau }_{n}^{\alpha
}i\omega _{n}^{\alpha }(\bar{w}_{n}^{\alpha }\dot{w}_{n}^{\alpha }-\dot{\bar{%
w}}_{n}^{\alpha }w_{n}^{\alpha }),  \label{uni1}
\end{equation}%
by (\ref{Lipfree}). Using the theorem (\ref{teo}), the universal kinetic
terms of the gauge sector can be written in the form 
\begin{equation}
\sum_{a}\sum_{n\in \mathcal{\hat{U}}}\mathbb{\tau }_{n}^{a}i\omega
_{n}^{a}\Delta (\bar{u}_{n}^{aT}\sigma _{1}\dot{u}_{n}^{a}-\dot{\bar{u}}%
_{n}^{aT}\sigma _{1}u_{n}^{a})=\sum_{a}\sum_{n\in \mathcal{\hat{U}}}\mathbb{%
\tau }_{n}^{a}i\omega _{n}^{a}(\bar{v}_{n}^{aT}\dot{u}_{n}^{a}-v_{n}^{aT}%
\dot{\bar{u}}_{n}^{a}+\dot{v}_{n}^{aT}\bar{u}_{n}^{a}-\dot{\bar{v}}%
_{n}^{aT}u_{n}^{a}),  \label{uni2}
\end{equation}%
the right-hand side being obtained using the properties (\ref{prope}) of
appendix \ref{DeltaProp}.

\subsection{Nontrivial boundary conditions}

\label{nontrbc}

So far, we have been working with trivial boundary conditions $\left. \phi
^{I}\right\vert _{\partial \Omega }=0$. Now we treat the case of general
Dirichlet boundary conditions%
\begin{equation}
\left. \phi ^{I}(t,\bm{x})\right\vert _{\partial \Omega }=f^{I}(t,\bm{x}%
_{\partial \Omega }),  \label{bondagen}
\end{equation}%
where $f^{I}$ are given functions, and $\bm{x}_{\partial \Omega }$ denotes
the space variables restricted to $\partial \Omega $. We want to show that
we can reduce this situation to the previous one, with few minor
modifications. In particular, the eigenfunctions, the frequencies and the
orthonormality relations remain the same.

First, we shift the fields $\phi ^{I}$ by some functions $\phi _{0}^{I}(t,%
\bm{x})$ that coincide with $f^{I}(t,\bm{x}_{\partial \Omega })$ on $%
\partial \Omega $: 
\begin{equation}
\phi ^{I}(t,\bm{x})=\phi _{0}^{I}(t,\bm{x})+\varphi ^{I}(t,\bm{x}),\qquad
\phi _{0}^{I}(t,\bm{x}_{\partial \Omega })=f^{I}(t,\bm{x}_{\partial \Omega
}),  \label{shifto}
\end{equation}%
so that the shifted fields vanish on $\Omega $:%
\begin{equation}
\left. \varphi ^{I}(t,\bm{x})\right\vert _{\partial \Omega }=0.
\label{bondagen0}
\end{equation}%
After the shift, we are free to integrate the space integrals by parts, to
move the space derivatives that act on any $\varphi ^{I}$ somewhere else.

By the assumptions we have made on the structure of the Lagrangian $L(\phi ,%
\dot{\phi})$, its expansion can be written as 
\begin{equation}
L(\phi ,\dot{\phi})=L_{0}+\varphi ^{I}A^{I}(\phi _{0})+\dot{\varphi}%
^{I}B^{I}(\phi _{0})+\nabla (\varphi ^{I}C^{I}(\phi _{0}))+L_{\varphi
}(\varphi ,\dot{\varphi})\equiv \tilde{L}_{\varphi }(\varphi ,\dot{\varphi}),
\label{lt}
\end{equation}%
where $L_{0}$ is $\varphi $-independent and $L_{\varphi }(\varphi ,\dot{%
\varphi})=L_{\text{free}}(\varphi ,\dot{\varphi})+$ interactions, by (\ref%
{Lfreeint}). We can ignore the $C$-term, since it disappears once we
integrate on the space manifold $\Omega $, by (\ref{bondagen0}). Were it
just for $L_{\varphi }(\varphi ,\dot{\varphi})$ (and $L_{0}$) we could apply
the formulation developed so far with no modifications. We want to explain
how to treat the corrections proportional to $A$ and $B$ (which need not be
perturbative).

Let us define 
\begin{equation}
\frac{\partial L_{\varphi }(\varphi ,\dot{\varphi})}{\partial \dot{\varphi}%
^{I}}=\pi _{\varphi }^{I}(\varphi ,\dot{\varphi}),\qquad \frac{\partial 
\tilde{L}_{\varphi }(\varphi ,\dot{\varphi})}{\partial \dot{\varphi}^{I}}=%
\tilde{\pi}_{\varphi }^{I}(\varphi ,\dot{\varphi})=\pi _{\varphi
}^{I}(\varphi ,\dot{\varphi})+B^{I},  \label{pis}
\end{equation}%
which invert to%
\begin{equation}
\dot{\varphi}^{I}=F^{I}(\varphi ,\pi _{\varphi }),\qquad \dot{\varphi}^{I}=%
\tilde{F}^{I}(\varphi ,\tilde{\pi}_{\varphi }),  \label{fid}
\end{equation}%
for certain functions $F^{I}$ and $\tilde{F}^{I}$. We have the Hamiltonians%
\begin{equation}
H_{\varphi }(\pi _{\varphi },\varphi )=\pi _{\varphi }^{I}F^{I}(\varphi ,\pi
_{\varphi })-L_{\varphi }(\varphi ,F(\varphi ,\pi _{\varphi })),\quad \tilde{%
H}_{\varphi }(\tilde{\pi}_{\varphi },\varphi )=\tilde{\pi}_{\varphi }^{I}%
\tilde{F}^{I}(\varphi ,\tilde{\pi}_{\varphi })-\tilde{L}_{\varphi }(\varphi ,%
\tilde{F}(\varphi ,\tilde{\pi}_{\varphi })),  \label{HH}
\end{equation}%
and the extended Lagrangians%
\begin{equation}
L_{\varphi }^{\prime }=\frac{1}{2}(\pi _{\varphi }^{I}\dot{\varphi}^{I}-\dot{%
\pi}_{\varphi }^{I}\varphi ^{I})-H_{\varphi }(\pi _{\varphi },\varphi
),\qquad \tilde{L}_{\varphi }^{\prime }=\frac{1}{2}(\tilde{\pi}_{\varphi
}^{I}\dot{\varphi}^{I}-\dot{\tilde{\pi}}_{\varphi }^{I}\varphi ^{I})-\tilde{H%
}_{\varphi }(\tilde{\pi}_{\varphi },\varphi ).  \label{LpLp}
\end{equation}%
Equating the two expressions of $\dot{\varphi}^{I}$ in (\ref{fid}) and using
the last identity of (\ref{pis}), we get%
\begin{equation*}
\tilde{F}^{I}(\varphi ,\tilde{\pi}_{\varphi })=F^{I}(\varphi ,\tilde{\pi}%
_{\varphi }-B).
\end{equation*}%
Using (\ref{lt}), (\ref{LpLp}) and (\ref{HH}), it is easy to work out the
difference%
\begin{eqnarray}
\Delta L_{\varphi }^{\prime } &\equiv &\tilde{L}_{\varphi }^{\prime }-\left.
L_{\varphi }^{\prime }\right\vert _{\pi _{\varphi }\rightarrow \tilde{\pi}%
_{\varphi }}=L_{0}+\varphi ^{I}A^{I}+\tilde{\pi}_{\varphi }^{I}\left[
F^{I}(\varphi ,\tilde{\pi}_{\varphi })-F^{I}(\varphi ,\tilde{\pi}_{\varphi
}-B)\right]  \notag \\
&&+B^{I}F^{I}(\varphi ,\tilde{\pi}_{\varphi }-B)+L_{\varphi }(\varphi
,F(\varphi ,\tilde{\pi}_{\varphi }-B))-L_{\varphi }(\varphi ,F(\varphi ,%
\tilde{\pi}_{\varphi })).  \label{diffe}
\end{eqnarray}%
If we switch the interactions off, we have $L_{\varphi }(\varphi ,\dot{%
\varphi})=L_{\text{free}}(\varphi ,\dot{\varphi})$, and the functions $%
F^{I}(\varphi ,\pi _{\varphi })$ become linear. Then formula (\ref{diffe})
tells us that $\Delta L_{\varphi }^{\prime }$ is made of linear terms, plus
interactions. In particular, the quadratic part of $\tilde{L}_{\varphi
}^{\prime }$ coincides with the quadratic part of $\left. L_{\varphi
}^{\prime }\right\vert _{\pi _{\varphi }\rightarrow \tilde{\pi}_{\varphi }}$.

At this point, we make the analogues of the shifts (\ref{pipi}), 
\begin{equation}
\bar{\tilde{\pi}}_{\varphi }^{I}=\tilde{\pi}_{\varphi }^{I}-\mathcal{B}%
^{IJi}(0)\partial _{i}\varphi ^{J}.  \label{pipi2}
\end{equation}%
They do not change the structure of $\Delta L_{\varphi }^{\prime }$, because
they do not involve time derivatives, and send linear terms into linear
terms, interaction terms into interaction terms. As far as the quadratic
part of $\tilde{L}_{\varphi }^{\prime }$ is concerned, it is equal to the
ones of (\ref{linva}) and (\ref{lip}) with the replacements $\phi
^{I}\rightarrow \varphi ^{I}$, $\bar{\pi}_{\phi }^{I}\rightarrow \bar{\tilde{%
\pi}}_{\varphi }^{I}$. Hence, if we expand the pair $\bar{\tilde{\pi}}%
_{\varphi }^{I},\varphi ^{I}$ exactly as we expanded $\bar{\pi}_{\phi
}^{I},\phi ^{I}$ before, we obtain the same quadratic part we had before, (%
\ref{Lip}) and (\ref{Lipfree}), plus interactions, plus linear terms (due to 
$\Delta L_{\varphi }^{\prime }$).

Note that $\bar{\tilde{\pi}}_{\varphi }^{I}$ vanishes on the boundary $%
\partial \Omega $ by construction, so to speak, since it is expanded in a
basis of functions that vanish there. Yet, we recall that no convergence
requirements are imposed on the expansion: the expansion itself must be
taken as the very definition of what $\left. \bar{\tilde{\pi}}_{\varphi
}^{I}\right\vert _{\partial \Omega }=0$ truly means. The same can be said of 
$\varphi $ and $\left. \varphi \right\vert _{\partial \Omega }=0$. As we
have already noted, the functional integral is defined by the very same
expansions.

As a result, we obtain a Lagrangian that has the same structure as before,
apart from including extra terms that are linear in the coherent states (and
terms that are independent of them). The endpoint corrections are
unmodified, because $\Delta L_{\varphi }^{\prime }$ does not contain time
derivatives. The complete action has the form (\ref{action}), plus the
corrections due to $\Delta L_{\varphi }^{\prime }$:%
\begin{eqnarray}
S &=&-i\sum_{I=1}^{r}\sum_{n\in \mathcal{\hat{U}}}\mathbb{\tau }%
_{n}^{I}\omega _{n}^{I}\left( \bar{z}_{n\text{f}}^{I}z_{n}^{I}(t_{\text{f}})+%
\bar{z}_{n}^{I}(t_{\text{i}})z_{n\text{i}}^{I}\right)  \notag \\
&&+\int_{t_{\text{i}}}^{t_{\text{f}}}\mathrm{d}t\left[ \mathcal{L}_{\text{%
free}}^{\prime }+\mathcal{L}_{\text{int}}^{\prime }+\sum_{I=1}^{r}\sum_{n\in 
\mathcal{\hat{U}}}\left( h_{n}^{I}z_{n}^{I}+h_{n}^{I\ast }\bar{z}%
_{n}^{I}\right) +k\right] ,  \label{complaction}
\end{eqnarray}%
for some, possibly time-dependent, functions $h_{n}^{I}$ and $k$.

As far as the local symmetries are concerned, the shift (\ref{shifto}) does
change the expressions of the transformations, unless the functions $\phi
_{0}^{I}$ are gauge invariant, which they must be, because we cannot build
physical quantities with unphysical boundary conditions. Referring to the
splitting of $\phi ^{I}$ into the three subsets $\phi ^{I_{+}}$, $\phi
^{I_{-}}$ and $\phi ^{I_{0}}$, we must require%
\begin{equation}
\phi _{0}^{I_{+}}=\phi _{0}^{I_{-}}=0.  \label{bgaugecond}
\end{equation}

Although we may sometimes relax the requirements (\ref{endrestr}) on the
initial and final conditions, we are definitely not going to relax the
requirements (\ref{bgaugecond}) on the boundary conditions, because there is
no reason to do so. Having specified this, we have proved that the situation
of general Dirichlet boundary conditions (\ref{bondagen}) reduces to the one
of vanishing boundary conditions, apart from some extra terms that are
linear in the coherent states, which are no source of worry.

\bigskip

In the case of gravity, we also need to extend the results to interaction
Lagrangians that contain arbitrarily many derivatives of the fields (as long
as their number grows together with the power of some coupling constant),
and show that we can rearrange the Lagrangian to have a final action with
the form and the properties of (\ref{complaction}). We deal with this aspect
in appendix \ref{hdinte}.

In conclusion, we have developed the general theory of coherent states for
local symmetries. It remains to use the results of \cite%
{AbsoPhys,AbsoPhysGrav} to arrange gauge invariance and general covariance
in the way we need. We do this in the next sections. Once that goal is
achieved, the results of this section, combined with those of \cite{MQ},
allow us to build the unitary evolution operator $U(t_{\text{f}},t_{\text{i}%
})$.

\section{Gauge theories: rearranging the Lagrangian}

\label{gaugetheories}\setcounter{equation}{0}

In gauge theories, we need to face a nontrivial issue: how can we specify
gauge invariant initial, final and boundary conditions? Giving the field
strength $F_{\mu \nu }$ is a possibility, but only in QED, because in
non-Abelian theories it is not gauge invariant. And even in QED, there
remains to give gauge invariant conditions for electrons.

These problems can be solved by introducing gauge invariant fields as
explained in ref. \cite{AbsoPhys,AbsoPhysGrav}. The goal is achieved by
means of a particular purely virtual extension of the theory. The physical
particles, the $S$ matrix amplitudes and the correlation functions of common
(nonlinear) composite fields (such as $F_{\mu \nu }^{a}F^{\mu \nu a}$, $\bar{%
\psi}\psi $, $\bar{\psi}\gamma ^{\mu }\psi $, etc.) do not change\footnote{%
In all such cases, the extension amounts to inserting \textquotedblleft
1\textquotedblright , written in a complicated way (a new type of
\textquotedblleft 1\textquotedblright\ with respect to the \textquotedblleft
1\textquotedblright\ used to gauge-fix the theory).}. Nevertheless, the
extension provides tools to define new, physical correlation functions, such
as the ones that contain insertions of gauge invariant fields, and calculate
them perturbatively. As we are going to show, it also allows us to specify
gauge invariant initial, final and boundary conditions at finite $\tau $ on
a compact $\Omega $.

The extension consists of a certain set of purely virtual extra fields. In
gauge theories \cite{AbsoPhys} we have scalar fields $\phi ^{a}$, together
with their anticommuting partners $\bar{H}^{a}$ and $H^{a}$, where $a$ is
the Lie-algebra index. In addition, it may be convenient to include certain
Lagrange multipliers $E^{a}$. The extension preserves renormalizability and
unitarity. Unitarity is also the reason why the extra fields must be purely
virtual: if not, the extension would propagate ghosts, and unitarity would
be lost.

The crucial property, for our purposes, is that the extension allows us to
switch to gauge invariant variables, and trivialize the gauge symmetry, to
fulfill the conditions (\ref{lincond}). The coherent states are then
introduced as explained in the previous section, and the rest follows from
there.

We focus on pure gauge theories, for simplicity, since there is no
difficulty to add the matter fields, when needed. We separate the time and
space components of the gauge fields by writing $A^{\mu }=(A^{0},\bm{A})$.
The dot on a field denotes its time derivative.

Instead of the common Lorenz gauge-fixing, given by the function $\partial
_{\mu }A^{\mu a}$, we use the more general function $\xi \dot{A}^{0a}+%
\bm{\nabla }\cdot \bm{A}^{a}$, where $\xi $ is an unspecified constant,
which allows us to interpolate between different gauge choices. Then, the
gauge-fixed Lagrangian is%
\begin{equation*}
\tilde{L}_{\text{gf}}=\frac{1}{2}\bm{F}^{a}\cdot \bm{F}^{a}-\frac{1}{4}%
F_{ij}^{a\hspace{0.01in}2}+B^{a}\left( \xi \partial _{0}A^{0a}+\bm{\nabla }%
\cdot \bm{A}^{a}+\frac{\lambda }{2}B^{a}\right) -\xi \bar{C}^{a}D_{0}\dot{C}%
^{a}+\bar{C}^{a}\bm{\nabla }\cdot \bm{D}C^{a},
\end{equation*}%
where $D_{\mu }=(D_{0},\bm{D})$ is the covariant derivative, and $\bm{F}^{a}=%
\bm{\dot{A}}^{a}+\bm{\nabla }A^{0a}+gf^{abc}A^{0b}\bm{A}^{c}$ are the 0$i$
components of the field strength. The so-called special gauge \cite%
{unitarityc}, which we use in the examples of sections \ref%
{semiinfinitecylinder} and \ref{finitecylinder}, is $\xi =\lambda $. The
Feynman gauge is $\xi =\lambda =1$. Like the Feynman gauge, the special
gauge allows us to simplify many formulas. In addition, it allows us to keep
a gauge-fixing parameter free, which is useful to study the gauge
independence of the physical quantities.

First, we rearrange $\tilde{L}_{\text{gf}}$, since no fields should be
differentiated twice. For reasons that will become clear later, we also turn
the derivatives contained in the gauge-fixing function onto $B$. We thus
obtain%
\begin{equation*}
L_{\text{gf}}=\frac{1}{2}\bm{F}^{a}\cdot \bm{F}^{a}-\frac{1}{4}F_{ij}^{a%
\hspace{0.01in}2}-\xi \dot{B}^{a}A^{0a}-\bm{\nabla }B^{a}\cdot \bm{A}^{a}+%
\frac{\lambda }{2}B^{a}B^{a}+\xi \dot{\bar{C}}^{a}D_{0}C^{a}-(\bm{\nabla }%
\bar{C}^{a})\cdot \bm{D}C^{a}.
\end{equation*}

Next, we introduce extra scalar fields $\phi ^{a}$, and their anticommuting
partners $H^{a}$, transforming as \cite{AbsoPhys}%
\begin{equation}
\delta _{\Lambda }\phi =\frac{ig\hspace{0.01in}\mathrm{ad}_{\phi }}{\mathrm{e%
}^{ig\hspace{0.01in}\mathrm{ad}_{\phi }}-1}\Lambda \equiv R^{a}(\phi
,\Lambda )T^{a},\qquad \delta _{\Lambda }H^{a}=H^{b}\frac{\delta R^{a}(\phi
,\Lambda )}{\delta \phi ^{b}}.  \label{localtra}
\end{equation}%
where $\phi =\phi ^{a}T^{a}$, $\Lambda =\Lambda ^{a}T^{a}$, $\Lambda ^{a}(x)$
are the parameters of the gauge transformation, $\mathrm{ad}_{\phi }X\equiv
\lbrack \phi ,X]$, and $T^{a}$ are the Lie algebra generators. We also
introduce gauge invariant antipartners $\bar{H}^{a}$ and Lagrange
multipliers $E^{a}$.

The gauge invariant fields $A_{\mu \text{d}}=A_{\mu \text{d}}^{a}T^{a}$ are
then%
\begin{equation}
A_{\mu \text{d}}\equiv \mathrm{e}^{-ig\hspace{0.01in}\mathrm{ad}_{\phi
}}A_{\mu }-\frac{1-\mathrm{e}^{-ig\hspace{0.01in}\mathrm{ad}_{\phi }}}{ig%
\hspace{0.01in}\mathrm{ad}_{\phi }}(\partial _{\mu }\phi ),\qquad \delta
_{\Lambda }A_{\mu \text{d}}=0,  \label{amud}
\end{equation}%
where the subscript \textquotedblleft d\textquotedblright\ stands for
\textquotedblleft dressed\textquotedblright .

The extension is a sort of mirror of the gauge-fixing sector. However, it
must be gauge invariant. In its most convenient (and manifestly power
counting renormalizable) form, it is specified by a function $\tilde{\xi}%
\dot{A}_{\text{d}}^{0a}+\bm{\nabla }\cdot \bm{A}_{\text{d}}^{a}$, where $%
\tilde{\xi}$ is a free constant. It reads%
\begin{equation*}
L_{\text{ext}}=E^{a}\left( \tilde{\xi}\dot{A}_{\text{d}}^{0a}+\bm{\nabla }%
\cdot \bm{A}_{\text{d}}^{a}+\frac{\tilde{\lambda}}{2}E^{a}\right) -\tilde{\xi%
}\dot{\bar{H}}^{a}\frac{\delta A_{\text{d}}^{0a}}{\delta \phi ^{b}}H^{b}-(%
\bm{\nabla }\bar{H}^{a})\cdot \frac{\delta \bm{A}_{\text{d}}^{a}}{\delta
\phi ^{b}}H^{b},
\end{equation*}%
where $\tilde{\lambda}$ is another free constant. This expression of $L_{%
\text{ext}}$ is already rearranged (with respect to the expression appearing
in \cite{AbsoPhys}) to eliminate the double derivatives. It is easy to check
that $L_{\text{ext}}$ is invariant under the local transformations (\ref%
{localtra}) (for details on this, see \cite{AbsoPhys}).

The total action is $L_{\text{tot}}=L_{\text{gf}}+L_{\text{ext}}$. The
parameters $\tilde{\xi}$ and $\tilde{\lambda}$ are part of the large
arbitrariness we have, when we want to dress the elementary fields and make
them gauge invariant. They are unique, however, in a power counting
renormalizable context (preserving invariance under space rotations).
Physically, they may parametrize different interplays between the physical
process and the external environment, or the experimental apparatus.

The extension is equal to \textquotedblleft 1\textquotedblright\ on standard
gauge invariant correlation functions (where \textquotedblleft
standard\textquotedblright\ means: independent of $\phi $, $\bar{H}$, $H$
and $E$), as well as on the $S$ matrix amplitudes, at $\tau =\infty $, $%
\Omega =\mathbb{R}^{3}$. We can prove this fact as follows. Focus on the $E$%
-dependent terms 
\begin{equation*}
E^{a}\left( V^{a}+\frac{\tilde{\lambda}}{2}E^{a}\right) ,\qquad V^{a}\equiv 
\tilde{\xi}\dot{A}_{\text{d}}^{0a}+\bm{\nabla }\cdot \bm{A}_{\text{d}}^{a}.
\end{equation*}%
Insert \textquotedblleft 1\textquotedblright\ in the form of the Gaussian
integral with Lagrangian $-\tilde{\lambda}(Q^{a}-E^{a})^{2}/2$, where $Q^{a}$
are extra integration variables. We have%
\begin{equation}
E^{a}\left( V^{a}+\frac{\tilde{\lambda}}{2}E^{a}\right) -\frac{\tilde{\lambda%
}}{2}(Q^{a}-E^{a})^{2}=E^{a}(V^{a}+\tilde{\lambda}Q^{a})-\frac{\tilde{\lambda%
}}{2}Q^{a}Q^{a}.  \label{last}
\end{equation}%
Next: $i$) integrate on $E^{a}$, which gives a functional delta function $%
\bm{\delta }$; $ii$) integrate on $\bar{H}^{a}$ and $H^{a}$, which gives a
functional determinant $\bm{J}$; $iii$) integrate on $\phi ^{a}$, which
appears only in $\bm{\delta }$ and $\bm{J}$; this integral gives 1, because $%
\bm{J}$ is there precisely for this purpose; finally, $iv$) integrate on $%
Q^{a}$, which also gives 1, since the only $Q^{a}$ dependence that survives
the first three operations is the one contained in the last term of (\ref%
{last}).

This chain of operations cannot be repeated as is when the insertions are $%
\phi $ dependent, as are those made of the invariant fields $A_{\mu \text{d}%
} $. Thus, the gauge invariant insertions built with $\phi $ provide new,
physical correlation functions and amplitudes. What we want to show is that
these properties also allow us to study amplitudes between arbitrary gauge
invariant initial and final states, with arbitrary gauge invariant boundary
conditions, in a finite interval of time $\tau $ and on a compact space
manifold $\Omega $.

One might object that the fields $\phi ^{a}$ become propagating, as well as $%
\bar{H}^{a}$ and $H^{a}$. What are these fields, physically? They might even
be ghosts, on general grounds. On top of that, we do not want to change the
theory. We just want to study less common features of a standard theory.

These are the reasons why the whole extension has to be purely virtual. The
extra fields $\phi ^{a}$, $\bar{H}^{a}$ and $H^{a}$ propagate ghosts if they
are treated as ordinary fields. They do not, if they are purely virtual. If
the whole extension is purely virtual, it does not inject new degrees of
freedom into the theory, and can be used as a mere mathematical tool to
study uncommon quantities of a common theory.

Another great advantage of the extension is that it allows us to
\textquotedblleft trivialize\textquotedblright\ the gauge symmetry, by
switching to appropriate dual variables. For example, we can abandon the
original gauge potential $A_{\mu }$ in favor of the gauge invariant one $%
A_{\mu \text{d}}$. We can also abandon the parameters $\Lambda $ of the
gauge transformation in favor of 
\begin{equation}
\Lambda _{\text{d}}^{a}\equiv R^{a}(\phi ,\Lambda )=\delta _{\Lambda }\phi
^{a}.  \label{gauga}
\end{equation}%
If we express the gauge symmetry this way, it becomes trivial: $\delta \phi
=\Lambda _{\text{d}}$, $\delta A_{\mu \text{d}}=0$. Then, we introduce new
Faddeev-Popov ghosts $C_{\text{d}}^{a}$, by means of the identification $%
\Lambda _{\text{d}}^{a}=\theta C_{\text{d}}^{a}$, where $\theta $ is a
constant, anticommuting parameter. Since the gauge symmetry is just an
arbitrary shift of $\phi ^{a}$, its closure is trivial, so we can take $%
\delta C_{\text{d}}^{a}=0$. We can define new, gauge invariant anticommuting
partners $H_{\text{d}}$ by means of the relations $H=R(-\phi ,H_{\text{d}})$.

Inverting (\ref{amud}), we can use the relations%
\begin{equation}
A_{\mu }=\mathrm{e}^{ig\hspace{0.01in}\mathrm{ad}_{\phi }}A_{\mu \text{d}}-%
\frac{1-\mathrm{e}^{ig\hspace{0.01in}\mathrm{ad}_{\phi }}}{ig\hspace{0.01in}%
\mathrm{ad}_{\phi }}(\partial _{\mu }\phi ),\qquad R(\phi ,C)=C_{\text{d}%
},\qquad H=R(-\phi ,H_{\text{d}}),  \label{switch}
\end{equation}%
as a change of variables in the functional integral, to switch from the
original variables $A_{\mu }$, $C$, $\phi $ and $H$ to\ the dual variables $%
A_{\mu \text{d}}$, $C_{\text{d}}$, $\phi $ and $H_{\text{d}}$. The switch
has a trivial Jacobian determinant (if we use the dimensional regularization 
\cite{dimreg}). We do not change $\bar{C}$, $B$, $\bar{H}$ and $E$.

It is much easier to specify gauge invariant initial, final and boundary
conditions by means of the dual variables. To make the notation lighter, we
put a tilde on $A_{\mu }^{a}$, $C^{a}$ and $H^{a}$, to emphasize that they
are functions of $A_{\mu \text{d}}^{a}$, $C_{\text{d}}^{a}$, $\phi ^{a}$ and 
$H_{\text{d}}^{a}$, now, and then drop the \textquotedblleft
d\textquotedblright\ in $A_{\mu \text{d}}^{a}$, $C_{\text{d}}^{a}$, $H_{%
\text{d}}^{a}$. It will be sufficient to recall that, in the new notation, $%
A_{\mu }^{a}$ is inert under the gauge transformations ($\delta _{\Lambda
}A_{\mu }^{a}=0$), and so are $C^{a}$ and $H^{a}$.

After the switch (\ref{switch}), the multipliers $E^{a}$ remain non
derivative (differently from $B$), so we integrate them out. At the end
(check \cite{AbsoPhys} for details), we obtain the total action 
\begin{eqnarray}
L_{\text{tot}} &=&\frac{1}{2}\bm{F}^{a}\cdot \bm{F}^{a}-\frac{1}{4}F_{ij}^{a%
\hspace{0.01in}2}-\xi \dot{B}^{a}\tilde{A}^{0a}-\bm{\nabla }B^{a}\cdot %
\bm{\tilde{A}}^{a}+\frac{\lambda }{2}(B^{a})^{2}-\frac{1}{2\tilde{\lambda}}%
\left( \tilde{\xi}\dot{A}^{0a}+\bm{\nabla }\cdot \bm{A}^{a}\right) ^{2} 
\notag \\
&&+\xi \dot{\bar{C}}^{a}\frac{\delta \tilde{A}^{0a}}{\delta \phi ^{b}}C^{b}+(%
\bm{\nabla }\bar{C}^{a})\cdot \frac{\delta \bm{\tilde{A}}^{a}}{\delta \phi
^{b}}C^{b}+\tilde{\xi}\dot{\bar{H}}^{a}D_{0}H^{a}-(\bm{\nabla }\bar{H}%
^{a})\cdot \bm{D}H^{a}.  \label{totala}
\end{eqnarray}

The trivialized local symmetry is%
\begin{equation}
\delta \phi =\Lambda =\theta C,\qquad \delta \bar{C}=\theta B,\qquad \delta
B=\delta C=\delta A_{\mu }=\delta \bar{H}=\delta H=0.  \label{gaugetransf}
\end{equation}%
Note that $L_{\text{tot}}$ is invariant without adding total derivatives.
Thus, we are in the conditions of section \ref{cohegauge}. We can define the
coherent states as explained there, and from there build the unitary
operator $U(t_{\text{f}},t_{\text{i}})$ as explained in \cite{MQ}.

\section{Gauge theories: quadratic sector}

\label{cohequad}\setcounter{equation}{0}

In this section we explain how to introduce coherent states in the
free-field limit of gauge theories, which is the key part of the problem. In
the next section it will be relatively straightforward to include the
interactions.

The quadratic part of the Lagrangian is practically the same as if we were
working in QED. Thus, we suppress the index $a$ and write, from (\ref{totala}%
), 
\begin{eqnarray}
L_{\text{tot}} &=&\frac{1}{2}(\bm{\dot{A}}+\bm{\nabla }A_{0})^{2}-\frac{1}{4}%
F_{ij}^{\hspace{0.01in}2}-\frac{1}{2\tilde{\lambda}}\left( \tilde{\xi}\dot{A}%
^{0}+\bm{\nabla }\cdot \bm{A}\right) ^{2}-\xi \dot{B}A^{0}-\bm{\nabla }%
B\cdot \bm{A}  \label{Ltot} \\
&&-\xi \dot{B}\dot{\phi}+\bm{\nabla }B\cdot \bm{\nabla }\phi +\frac{\lambda 
}{2}B^{2}+\xi \dot{\bar{C}}\dot{C}-\bm{\nabla }\bar{C}\cdot \bm{\nabla }C+%
\tilde{\xi}\dot{\bar{H}}\dot{H}-\bm{\nabla }\bar{H}\cdot \bm{\nabla }H+%
\mathcal{O}(g).  \notag
\end{eqnarray}

With the variables we have chosen, gauge invariance simply means invariance
under the transformations $\delta _{\Lambda }\phi =\Lambda =\theta C$, $%
\delta _{\Lambda }\bar{C}=\theta B$, all the other fields being inert. Note
that the first line of (\ref{Ltot}) is manifestly invariant, to the lowest
order, while the terms appearing in the second line transform into one
another, apart from the $H$-dependent ones, which are also invariant. Thus,
we are in the conditions of section \ref{cohegauge}.

The field variables are $\hat{\Phi}\equiv (\phi ,B,A^{0},\bm{A},C,\bar{C},H,%
\bar{H})$. From the moment, we ignore $H$ and $\bar{H}$, and restrict to $%
\tilde{\Phi}\equiv (\phi ,B,A^{0},\bm{A},C,\bar{C})$, because it is
straightforward to treat $H$ and $\bar{H}$ along the lines of ref. \cite{MQ}%
. We discuss them anyway in the next section, when we include the
interactions. For the time being, we also drop $\mathcal{O}(g)$.

The boundary conditions read%
\begin{equation}
\left. \tilde{\Phi}\right\vert _{_{\partial \Omega }}=(\phi
_{0},B_{0},A_{0}^{0},\bm{A}_{0},C_{0},\bar{C}_{0}),  \label{bounda}
\end{equation}%
where the list on the right-hand side collects given functions on $\partial
\Omega $. We can turn to vanishing boundary conditions by means of shifts 
\begin{equation}
\tilde{\Phi}\rightarrow \tilde{\Phi}+\tilde{\Phi}_{0},  \label{shift}
\end{equation}%
where $\tilde{\Phi}_{0}$ are functions defined on the whole of $\Omega $,
which coincide with the right-hand side of (\ref{bounda}) on $\partial
\Omega $. This way, the new $\tilde{\Phi}$ vanish on $\partial \Omega $.
Since the shift does not change the quadratic sector of the free Lagrangian,
on which we are concentrating in the present section, we take $\left. \tilde{%
\Phi}\right\vert _{\partial \Omega }=\tilde{\Phi}_{0}=0$ for the moment, and
leave the rest of the discussion\ to the next section. Note that $\left. 
\tilde{\Phi}\right\vert _{\partial \Omega }=0$ allows us to freely integrate
the space integrals by parts.

The momenta, which are%
\begin{eqnarray}
\pi _{\phi } &=&-\xi \dot{B},\qquad \pi _{B}=-\xi (\dot{\phi}+A^{0}),\qquad
\pi _{A^{0}}=-\frac{\tilde{\xi}}{\tilde{\lambda}}(\tilde{\xi}\dot{A}^{0}+%
\bm{\nabla }\cdot \bm{A)},  \notag \\
\bm{\pi }_{A} &=&\bm{\dot{A}}+\bm{\nabla }A^{0},\qquad \pi _{\bar{C}}=\xi 
\dot{\bar{C}},\qquad \pi _{C}=\xi \dot{C},  \label{moma}
\end{eqnarray}%
are either gauge invariant, or transform into one another:%
\begin{equation}
\delta _{\Lambda }\pi _{B}=-\theta \pi _{C},\qquad \delta _{\Lambda }\pi _{%
\bar{C}}=-\theta \pi _{\phi },\qquad \delta _{\Lambda }\pi _{\phi }=\delta
_{\Lambda }\pi _{A_{0}}=\delta _{\Lambda }\bm{\pi }_{A}=\delta _{\Lambda
}\pi _{C}=0.  \label{gauga2}
\end{equation}%
The Hamiltonian is $H=H_{\text{bos}}+H_{\text{gh}}$, where%
\begin{eqnarray*}
H_{\text{bos}} &=&-\frac{1}{\xi }\pi _{\phi }(\pi _{B}+\xi A^{0})-\frac{1}{2%
\tilde{\xi}}\pi _{A^{0}}\left( \frac{\tilde{\lambda}}{\tilde{\xi}}\pi
_{A^{0}}+2\bm{\nabla }\cdot \bm{A}\right) +\frac{1}{2}\bm{\pi }_{A}(\bm{\pi }%
_{A}-2\bm{\nabla }A^{0}) \\
&&+\frac{1}{4}F_{ij}^{2}-\bm{\nabla }B\cdot \bm{\nabla }\phi +\bm{\nabla }%
B\cdot \bm{A}-\frac{\lambda }{2}B^{2},\qquad \qquad H_{\text{gh}}=\frac{1}{%
\xi }\pi _{\bar{C}}\pi _{C}+\bm{\nabla }\bar{C}\cdot \bm{\nabla }C,
\end{eqnarray*}%
so the extended Lagrangian $L^{\prime }$ of formula (\ref{Lpcohe}) is 
\begin{equation}
L^{\prime }=L_{\text{bos}}^{\prime }+L_{\text{gh}}^{\prime },\quad L_{\text{%
bos}}^{\prime }=\frac{1}{2}(\pi _{\Phi }\dot{\Phi}-\dot{\pi}_{\Phi }\Phi
)-H_{\text{bos}},\qquad L_{\text{gh}}^{\prime }=\frac{1}{2}(\pi _{\bar{C}}%
\dot{C}-\dot{\pi}_{\bar{C}}C+\dot{\bar{C}}\pi _{C}-\bar{C}\dot{\pi}_{C})-H_{%
\text{gh}},  \label{Lbos}
\end{equation}%
where $\Phi =(\phi ,B,A^{0},\bm{A})$.

As explained in the previous two sections, it is convenient to introduce the
shifted momenta (\ref{pipi}), or (\ref{pipi2}), which are%
\begin{equation}
\bar{\pi}_{A_{0}}=\pi _{A_{0}}+\frac{\tilde{\xi}}{\tilde{\lambda}}\bm{\nabla
}\cdot \bm{A}=-\frac{\tilde{\xi}^{2}}{\tilde{\lambda}}\dot{A}^{0},\qquad %
\bm{\bar{\pi}}_{A}=\bm{\pi }_{A}-\bm{\nabla }A^{0}=\bm{\dot{A}},
\label{momat}
\end{equation}%
while the other momenta are unchanged. Defining $\bar{\Pi}=(\pi _{\phi },\pi
_{B},\bar{\pi}_{A_{0}},\bm{\bar{\pi}}_{A})$, the general form of the
Lagrangian $L_{\text{bos}}^{\prime }$ is 
\begin{equation*}
L_{\text{bos}}^{\prime }=\frac{1}{2}\left( 
\begin{tabular}{ll}
$\bar{\Pi}$ & $\Phi $%
\end{tabular}%
\right) \left( 
\begin{tabular}{cc}
$M$ & $K_{2}+\partial _{t}$ \\ 
$K_{2}^{T}-\partial _{t}$ & $N$%
\end{tabular}%
\right) \left( 
\begin{tabular}{l}
$\bar{\Pi}$ \\ 
$\Phi $%
\end{tabular}%
\right) ,\qquad M=\left( 
\begin{tabular}{cccc}
$0$ & $\frac{1}{\xi }$ & $0$ & $0$ \\ 
$\frac{1}{\xi }$ & $0$ & $0$ & $0$ \\ 
$0$ & $0$ & $\frac{\tilde{\lambda}}{\tilde{\xi}^{2}}$ & $0$ \\ 
$0$ & $0$ & $0$ & $-1$%
\end{tabular}%
\right) ,
\end{equation*}%
where $K_{2}$ is a constant matrix and $K_{2}^{T}$ is its transpose, while $%
N=N_{1}^{ij}\partial _{i}\partial _{j}+N_{2}^{i}\partial _{i}\partial
_{t}+N_{3}^{i}\partial _{i}+N_{4}$, where $N_{1}^{ij}$, $N_{2}^{i}$, $%
N_{3}^{i}$ and $N_{4}$ are other constant matrices. We do not specify them
here (and, besides, most of their entries are just zero, as in $M$), since
they can be read directly from (\ref{Lbos}). It is sufficient to note that $%
N_{1}^{ij}$, $N_{2}^{i}$ and $N_{4}$ are symmetric, while $N_{3}^{i}$ are
antisymmetric.

Ultimately, we are in the situation described in general terms in subsection %
\ref{freque}. We have eigenfunctions $\bar{\Pi}_{n}$, $\Phi _{n}$, with
(real) frequencies $\omega _{n}$, where $n$ is some label ranging in some
set $\mathcal{V}$. The complex conjugate eigenfunctions are those with some
\textquotedblleft conjugate\textquotedblright\ label $n^{\ast }$, i.e.,%
\begin{equation*}
\bar{\Pi}_{n}^{\ast }(\bm{x})=\bar{\Pi}_{n^{\ast }}(\bm{x}),\qquad \Phi
_{n}^{\ast }(\bm{x})=\Phi _{n^{\ast }}(\bm{x}),\qquad \omega _{n^{\ast
}}=-\omega _{n}.
\end{equation*}%
We then expand $\bar{\Pi}$ and $\Phi $ in such a basis:%
\begin{equation}
\left( 
\begin{tabular}{l}
$\bar{\Pi}$ \\ 
$\Phi $%
\end{tabular}%
\right) =\sum_{n\in \mathcal{V}}a_{n}\left( 
\begin{tabular}{l}
$\bar{\Pi}_{n}$ \\ 
$\Phi _{n}$%
\end{tabular}%
\right) ,  \label{expansion}
\end{equation}%
with $a_{n^{\ast }}=a_{n}^{\ast }$. As before, we write $\mathcal{V=U}\cup 
\mathcal{U}^{\ast }$, so that each pair $n,n^{\ast }$ is split between $%
\mathcal{U}$ and $\mathcal{U}^{\ast }$. The orthonormality\ relations are (%
\ref{ortono}). Using them, we can invert (\ref{expansion}) as in (\ref%
{coeffa}), and obtain the expansion of the integrated bosonic Lagrangian,
which reads 
\begin{equation}
\mathcal{L}_{\text{bos}}^{\prime }\equiv \int_{\Omega }L_{\text{bos}%
}^{\prime }\mathrm{d}^{3}\bm{x}=\sum_{n\in \mathcal{U}}i\mathbb{\tau }%
_{n}\omega _{n}(a_{n}^{\ast }\dot{a}_{n}-\dot{a}_{n}^{\ast
}a_{n})-2\sum_{n\in \mathcal{U}}\mathbb{\tau }_{n}\omega _{n}^{2}a_{n}^{\ast
}a_{n}.  \label{LnF}
\end{equation}

Since we have six independent fields (for every value of the Lie algebra
index $a$), which are the components $\phi $, $B$ and $A_{\mu }$ of $\Phi $,
we can distinguish six classes of frequencies $\omega _{n}$. Two of them,
which we denote by $\omega _{n}^{\text{g}}$ and $\omega _{n}^{\text{g\hspace{%
0.01in}}\prime }$, may depend on the gauge-fixing parameters $\xi $ and $%
\lambda $, while the other four may depend on the parameters $\tilde{\xi}$
and $\tilde{\lambda}$, but not on $\xi $ and $\lambda $.

Out of the four gauge independent frequencies, two are physical, denoted by $%
\omega _{n}^{\text{ph}}$ and $\omega _{n}^{\text{ph\hspace{0.01in}}\prime }$%
, and two must be quantized as purely virtual, denoted by $\omega _{n}^{%
\text{d}}$ and $\omega _{n}^{\text{d\hspace{0.01in}}\prime }$.

The distinction between the two classes of gauge independent frequencies is
somewhat flexible. In the absence of data (which require to make experiments
about scattering processes where the restrictions to finite $\tau $ and
compact $\Omega $ play crucial roles), the only theoretical constraints we
have are that: $a$) the eigenfunctions $\bar{\Pi}_{n}$, $\Phi _{n}$,
associated with each set of frequencies $\omega _{n}^{\text{g}}$, $\omega
_{n}^{\text{g\hspace{0.01in}}\prime }$, $\omega _{n}^{\text{ph}}$, $\omega
_{n}^{\text{ph\hspace{0.01in}}\prime }$, $\omega _{n}^{\text{d}}$ and $%
\omega _{n}^{\text{d\hspace{0.01in}}\prime }$, make a complete set for some
component of $\bar{\Pi}$, $\Phi $; $b$) altogether, they are a complete set
for $\bar{\Pi}$, $\Phi $; $c$) the eigenfunctions have the right limits for $%
\Omega \rightarrow \mathbb{R}^{3}$. Such limits are $\tilde{\xi}$ and $%
\tilde{\lambda}$ independent for $\omega _{n}^{\text{ph}}$, $\omega _{n}^{%
\text{ph\hspace{0.01in}}\prime }$, $\tilde{\xi}$ and $\tilde{\lambda}$
dependent for $\omega _{n}^{\text{d}}$ and $\omega _{n}^{\text{d\hspace{%
0.01in}}\prime }$.

\medskip

As an example of the flexibility we are referring to, we can consider linear
combinations of solutions whose frequencies have the same limits for $\Omega
\rightarrow \mathbb{R}^{3}$. As we show in the examples of sections \ref%
{semiinfinitecylinder} and \ref{finitecylinder}, if the relative
coefficients are appropriately oscillating, the mixing disappears when $%
\Omega \rightarrow \mathbb{R}^{3}$. This ambiguity reflects the large
arbitrariness we have, when we formulate quantum field theory in a finite
interval of time $\tau $, on a compact space manifold $\Omega $. Like the
parameters $\tilde{\xi}$ and $\tilde{\lambda}$, different choices of the
basis (\ref{Zn3}) may parametrize, in a way that remains to be clarified,
different interplays between the physical process we want to study and the
external environment where it is placed, or the apparatus we use to make the
measurements.

\medskip

In several cases, it may be helpful to first set $\tilde{\xi}=\tilde{\lambda}%
=1$, where the frequencies and eigenfunctions simplify and can often be
written explicitly, make the choices of basis there, and then extend the
choices to $\tilde{\xi},\tilde{\lambda}\neq 1$ by expanding in powers of $%
\delta _{\tilde{\xi}}=(\tilde{\xi}-1)/2$ and $\delta _{\tilde{\lambda}}=(%
\tilde{\lambda}-1)/2$.

\medskip

The gauge dependent frequencies $\omega _{n}^{\text{g}}$ and $\omega _{n}^{%
\text{g\hspace{0.01in}}\prime }$ can be quantized as purely virtual or not,
provided we implement this choice consistently everywhere. The physical
quantities are unaffected by the choice, because they are gauge independent.

\medskip

Writing $\mathcal{U=U}_{\text{g}}\mathcal{\cup U}_{\text{g}}^{\prime }%
\mathcal{\cup U}_{\text{d}}\mathcal{\cup U}_{\text{d}}^{\prime }\mathcal{%
\cup U}_{\text{ph}}\mathcal{\cup U}_{\text{ph}}^{\prime }\equiv \mathcal{%
\cup }_{I=1}^{6}\mathcal{U}^{I}$, the expansion (\ref{expansion}) becomes%
\begin{equation}
\left( 
\begin{tabular}{l}
$\bar{\Pi}$ \\ 
$\Phi $%
\end{tabular}%
\right) =\sum_{I=1}^{6}\sum_{n\in \mathcal{U}^{I}}\left[ \left( 
\begin{tabular}{l}
$\bar{\Pi}_{n}^{I}$ \\ 
$\Phi _{n}^{I}$%
\end{tabular}%
\right) z_{n}^{I}+\left( 
\begin{tabular}{l}
$\bar{\Pi}_{n}^{I\ast }$ \\ 
$\Phi _{n}^{I\ast }$%
\end{tabular}%
\right) \bar{z}_{n}^{I}\right] .  \label{expat}
\end{equation}
The coefficients are the coherent states\ 
\begin{equation}
Z_{n}(t)=(z_{n}^{\text{g}},z_{n}^{\text{g\hspace{0.01in}}\prime },z_{n}^{%
\text{d}},z_{n}^{\text{d\hspace{0.01in}}\prime },z_{n}^{\text{ph}},z_{n}^{%
\text{ph\hspace{0.01in}}\prime })=(z_{n}^{I})=(a_{n}).  \label{Zn}
\end{equation}%
The bosonic Lagrangian $\mathcal{L}_{\text{bos}}^{\prime }$ can be split
accordingly.

\medskip

The two gauge dependent frequencies $\omega _{n}^{\text{g}}$ and $\omega
_{n}^{\text{g\hspace{0.01in}}\prime }$ are easy to calculate, since they
must correspond, by the gauge symmetry, to those of the ghost Lagrangian $L_{%
\text{gh}}^{\prime }$. Repeating the procedure described above for $L_{\text{%
gh}}^{\prime }$, we find that the eigenfunctions we are talking about solve
the standard problem%
\begin{equation*}
\Delta C_{n}(\bm{x})=-\xi \omega _{n}^{2}C_{n}(\bm{x})\text{ in }\Omega
,\qquad \left. C_{n}\right\vert _{\partial \Omega }=0,
\end{equation*}%
and come in two copies (ghosts and antighosts).

\medskip

The gauge transformations of the coherent states can be derived from the
ones of the fields and the momenta, combined with the expansion (\ref{expat}%
), as explained in subsection \ref{gtcohe}. Since $\delta \phi =\theta C$, $%
\delta \bar{C}=\theta B$, there must be $\phi $ modes that transform into
the ghost ones, and antighost modes that transform into the $B$ ones. This
means that the $\phi $, $B$, $C$ and $\bar{C}$ expansions have the
structures 
\begin{eqnarray}
\left( 
\begin{tabular}{l}
$\phi $ \\ 
$B$%
\end{tabular}%
\right) &=&\left( 
\begin{tabular}{l}
$\phi _{n}$ \\ 
$0$%
\end{tabular}%
\right) z_{\phi n}+\left( 
\begin{tabular}{l}
$\phi _{n}^{\ast }$ \\ 
$0$%
\end{tabular}%
\right) \bar{z}_{\phi n}+\left( 
\begin{tabular}{l}
$\psi _{n}$ \\ 
$B_{n}$%
\end{tabular}%
\right) z_{Bn}+\left( 
\begin{tabular}{l}
$\psi _{n}^{\ast }$ \\ 
$B_{n}^{\ast }$%
\end{tabular}%
\right) \bar{z}_{Bn}+\cdots  \notag \\
\left( 
\begin{tabular}{l}
$C$ \\ 
$\bar{C}$%
\end{tabular}%
\right) &=&\left( 
\begin{tabular}{l}
$\phi _{n}$ \\ 
$\psi _{n}^{\prime }$%
\end{tabular}%
\right) z_{Cn}+\left( 
\begin{tabular}{l}
$\phi _{n}^{\ast }$ \\ 
$\psi _{n}^{\prime \ast }$%
\end{tabular}%
\right) \bar{z}_{Cn}+\left( 
\begin{tabular}{l}
$0$ \\ 
$B_{n}$%
\end{tabular}%
\right) z_{\bar{C}n}+\left( 
\begin{tabular}{l}
$0$ \\ 
$B_{n}^{\ast }$%
\end{tabular}%
\right) \bar{z}_{\bar{C}n}.  \label{bcc}
\end{eqnarray}%
where the sums on $n$ are understood, the dots collect the contributions of
the $A^{0}$ and $\bm{A}$ modes, and $\psi _{n}$ and $\psi _{n}^{\prime }$
are unspecified functions. The coherent states denoted by $z_{\phi n}$ and $%
\bar{z}_{\phi n}$ do not contribute to the expansions of $A^{0}$ and $\bm{A}$%
; the $A^{0}$, $\bm{A}$ modes may contribute to the expansion of $\phi $,
but not to the one of $B$.

The only nontrivial gauge transformations of the coherent states are%
\begin{equation}
\delta z_{\phi n}=\theta z_{Cn},\qquad \delta \bar{z}_{\phi n}=\theta \bar{z}%
_{Cn},\qquad \delta z_{\bar{C}n}=\theta z_{Bn},\qquad \delta \bar{z}_{\bar{C}%
n}=\theta \bar{z}_{Bn},  \label{cotra}
\end{equation}%
and the $\phi BC\bar{C}$ sector of the Lagrangian reads%
\begin{eqnarray}
L_{\phi BC\bar{C}}^{\prime } &=&-\sum_{n}i\omega _{n}(\bar{z}_{Bn}\dot{z}%
_{\phi n}-\dot{\bar{z}}_{Bn}z_{\phi n}+\dot{z}_{Bn}\bar{z}_{\phi n}-z_{Bn}%
\dot{\bar{z}}_{\phi n})+2\sum_{n}\omega _{n}^{2}(\bar{z}_{Bn}z_{\phi
n}+z_{Bn}\bar{z}_{\phi n})  \notag \\
&&\!\!\!\!\!\!\!\!\!\!\!\!{+\sum_{n}i\omega _{n}(\bar{z}_{\bar{C}n}\dot{z}%
_{Cn}-\dot{\bar{z}}_{\bar{C}n}z_{Cn}+\dot{z}_{\bar{C}n}\bar{z}_{Cn}-z_{\bar{C%
}n}\dot{\bar{z}}_{Cn})-2\sum_{n}\omega _{n}^{2}(\bar{z}_{\bar{C}n}z_{Cn}+z_{%
\bar{C}n}\bar{z}_{Cn}).\qquad}  \label{Lpgh}
\end{eqnarray}

\section{Gauge theories: interactions}

\label{inter}\setcounter{equation}{0}

Now that we have taken care of the quadratic part, we are ready to include
the interactions. Working out the momenta $\pi _{\Phi }$ from the Lagrangian
(\ref{totala}), we obtain 
\begin{eqnarray*}
\pi _{B}^{a} &=&-\xi \tilde{A}^{0a},\qquad \pi _{A^{0}}^{a}=-\frac{\tilde{\xi%
}^{2}}{\tilde{\lambda}}\dot{A}^{0}-\frac{\tilde{\xi}}{\tilde{\lambda}}%
\bm{\nabla }\cdot \bm{A},\qquad \bm{\pi }_{A}^{a}=\bm{F}^{a}, \\
\pi _{\bar{C}} &=&\xi \frac{1-\mathrm{e}^{-ig\hspace{0.01in}\mathrm{ad}%
_{\phi }}}{ig\hspace{0.01in}\mathrm{ad}_{\phi }}\dot{\bar{C}},\qquad \pi
_{C}^{a}=\xi \frac{\delta \tilde{A}^{0a}}{\delta \phi ^{b}}C^{b},\qquad \pi
_{\bar{H}}^{a}=\tilde{\xi}\dot{\bar{H}}^{a},\qquad \pi _{H}^{a}=\tilde{\xi}%
D_{0}H^{a},
\end{eqnarray*}%
plus $\pi _{\phi }$, which we do not report here, because its expression can
be read from the gauge transformations, which, by (\ref{deltap}), are still (%
\ref{gaugetransf}) and (\ref{gauga2}): $\theta \pi _{\phi }=-\delta \pi _{%
\bar{C}}$.

Then, we make the redefinition (\ref{pipi}). The only changes are 
\begin{equation}
\bar{\pi}_{A_{0}}^{a}=\pi _{A_{0}}^{a}+\frac{\tilde{\xi}}{\tilde{\lambda}}%
\bm{\nabla }\cdot \bm{A}^{a}\bm{=}-\frac{\tilde{\xi}^{2}}{\tilde{\lambda}}%
\dot{A}^{0},\qquad \bm{\bar{\pi}}_{A}^{a}=\bm{\pi }_{A}^{a}-\bm{\nabla }%
A^{0a}=\bm{\dot{A}}^{a}+gf^{abc}A^{0b}\bm{A}^{c}.  \label{pipi2g}
\end{equation}%
Since the differences between $\pi _{\hat{\Phi}}$ and $\bar{\pi}_{\hat{\Phi}%
} $ are gauge invariant, the gauge transformations of the new variables $%
\bar{\pi}_{\hat{\Phi}}$ and $\hat{\Phi}$ are simply%
\begin{equation}
\delta \phi =\Lambda =\theta C,\qquad \delta \bar{C}=\theta B,\qquad \delta 
\bar{\pi}_{B}=-\theta \bar{\pi}_{C},\qquad \delta \bar{\pi}_{\bar{C}%
}=-\theta \bar{\pi}_{\phi },  \label{cocc}
\end{equation}%
the other fields $\bar{\pi}_{\hat{\Phi}}$ and $\hat{\Phi}$ being invariant.

The case of trivial boundary conditions $\left. \hat{\Phi}\right\vert
_{_{\partial \Omega }}=0$, which are evidently gauge invariant, can be
treated straightforwardly. The quadratic Lagrangian, which defines the
coherent states and the expansions of the fields, is the one of the previous
section. The interacting sector $\mathcal{L}_{\text{int}}^{\prime }$ of the
total action (\ref{action}) can be easily expressed in terms of coherent
states, since it does not depend on their time derivatives.

The most general boundary conditions are $\left. \hat{\Phi}\right\vert
_{_{\partial \Omega }}=\hat{f}$, where $\hat{f}$ is a row of given functions
on $\partial \Omega $. To build physical amplitudes, we must choose a gauge
invariant and gauge nontrivial $\hat{f}$, which means set $\phi =B=C=\bar{C}%
=0$ on $\partial \Omega $. Since there are no theoretical or practical
motivations to relax these requirements, from now on we adopt the boundary
conditions%
\begin{equation}
\left. \hat{\Phi}\right\vert _{_{\partial \Omega }}=(0,0,A_{0}^{0},\bm{A}%
_{0},0,0,H_{0},\bar{H}_{0}).  \label{boundacy}
\end{equation}%
Then we make the shifts 
\begin{equation}
\hat{\Phi}\rightarrow \hat{\Phi}+\hat{\Phi}_{0},  \label{shiftc}
\end{equation}%
where $\hat{\Phi}_{0}$ are functions defined on the whole of $\Omega $, with
the sole requirement that they coincide with the right-hand side of (\ref%
{boundacy}) on $\partial \Omega $. After the shift, the boundary conditions
are $\left. \hat{\Phi}\right\vert _{_{\partial \Omega }}=0$, the gauge
transformations are still (\ref{gaugetransf}), and we can freely integrate
the space integrals by parts, to move space derivatives away from any field.

It is important to stress that the conditions (\ref{boundacy}) apply to the
Lagrangian (\ref{totala}), before even talking about momenta, so we do not
have to worry about the behaviors of the momenta on $\partial \Omega $ at
this stage.

Take the Lagrangian (\ref{totala}), and denote it by $L_{\text{tot}}(\hat{%
\Phi})=L_{\text{free}}(\hat{\Phi})+L_{\text{int}}(\hat{\Phi})$. Once we
implement the shift (\ref{shiftc}) on it, we obtain%
\begin{equation}
L_{\text{tot}}(\hat{\Phi}+\hat{\Phi}_{0})=L_{\text{tot}}(\hat{\Phi}_{0})+A(%
\hat{\Phi}_{0})\hat{\Phi}+B(\hat{\Phi}_{0})\dot{\hat{\Phi}}+\nabla (\hat{\Phi%
}C(\hat{\Phi}_{0}))+L(\hat{\Phi},\hat{\Phi}_{0}),  \label{lili}
\end{equation}%
where $L(\hat{\Phi},\hat{\Phi}_{0})=L_{\text{free}}(\hat{\Phi})+$
interactions. We can ignore the term $\nabla (\hat{\Phi}C(\hat{\Phi}_{0}))$,
since it disappears as soon as we integrate on the space manifold $\Omega $.
The quadratic sector of $L_{\text{tot}}(\hat{\Phi}+\hat{\Phi}_{0})$
coincides with $L_{\text{free}}(\hat{\Phi})$, which is the one of $L_{\text{%
tot}}(\hat{\Phi})$, up to interactions.

Next, we proceed as explained in subsection \ref{nontrbc}. We define the
momenta, redefine them according to (\ref{pipi2}) (that is to say, according
to (\ref{pipi2g}) with $\pi \rightarrow \tilde{\pi}$, $\bar{\pi}\rightarrow 
\bar{\tilde{\pi}}$), and get to the extended Lagrangian $\tilde{L}_{\varphi
}^{\prime }$. Since the quadratic sector of (\ref{lili}) is $L_{\text{free}}(%
\hat{\Phi})$, the eigenfunctions coincide with those we had with trivial
boundary conditions. So do the expansions in terms of coherent states (\ref%
{Zn}). Once we integrate the Lagrangian and include endpoint corrections, to
have the correct variational problem, the final action is (\ref{complaction}%
), which just contains some linear corrections (and possibly different
interactions) with respect to the action (\ref{action}).

Once we have the action, the theory can be phrased diagrammatically. The
diagrams are of the usual type, apart from the presence of external sources
and the discretizations of the loop momenta \cite{MQ}.

When we want a transition amplitude, we must choose initial and final
conditions $z_{n}^{I}(t_{\text{i}})=z_{n\text{i}}^{I}$, $\bar{z}_{n}^{I}(t_{%
\text{f}})=\bar{z}_{n\text{f}}^{I}$ for the coherent states. The physical
degrees of freedom are the transverse components of $\bm{A}$, which must be
quantized as physical particles. Their initial and final conditions $z_{n%
\text{i}}^{\text{ph}}$, $z_{n\text{i}}^{\text{ph\hspace{0.01in}}\prime }$, $%
\bar{z}_{n\text{f}}^{\text{ph}}$ and $\bar{z}_{n\text{f}}^{\text{ph\hspace{%
0.01in}}\prime }$ are free.

The gauge degrees of freedom are $\phi $, $B$, $C$ and $\bar{C}$. They can
be quantized as purely virtual or not, provided the choice is implemented
consistently everywhere. Their initial and final conditions are trivial,
i.e., 
\begin{equation}
z_{n\text{i}}^{\text{g}}=z_{n\text{i}}^{\text{g\hspace{0.01in}}\prime }=\bar{%
z}_{n\text{f}}^{\text{g}}=\bar{z}_{n\text{f}}^{\text{g\hspace{0.01in}}\prime
}=0,  \label{gcond}
\end{equation}%
and similarly for $C$ and $\bar{C}$.

The purely virtual fields are $A^{0}$, $H$, $\bar{H}$ and the longitudinal
components of $\bm{A}$. They are quantized as purely virtual particles, by
removing their on-shell contributions to the diagrams perturbatively to all
orders, according to the rules of ref.s \cite{diagrammarMio,PVP20}, and
setting the initial and final conditions of the coherent states associated
with them to zero. This means%
\begin{equation}
z_{n\text{i}}^{\text{d}}=z_{n\text{i}}^{\text{d\hspace{0.01in}}\prime }=\bar{%
z}_{n\text{f}}^{\text{d}}=\bar{z}_{n\text{f}}^{\text{d\hspace{0.01in}}\prime
}=0,  \label{pvcond}
\end{equation}%
and similarly for $H$ and $\bar{H}$.

The decomposition of $\bm{A}$ into \textquotedblleft
transverse\textquotedblright\ and \textquotedblleft
longitudinal\textquotedblright\ components is defined by the arrangement (%
\ref{expat}), after identifying the (physical vs purely virtual)
eigenfunctions (\ref{Zn})\ and their frequencies. We illustrate these facts
in the examples of the next two sections.

Note that we do not need to disentangle the physical and purely virtual
degrees of freedom on $\partial \Omega $, because purely virtual particles
are not required to have trivial boundary conditions \cite{MQ}. The freedom
associated with their boundary conditions may describe some sort of
interaction between the observer, or the environment, and the physical
process we are observing.

The unitarity equation $U^{\dag }U=1$ holds under appropriate assumptions
(such as the cancellation of the gauge anomalies at one loop). An easy way
to prove the statement is to formulate the gauge sector (identified by the
fields $\phi $, $B$, $C$ and $\bar{C}$) as purely virtual, as in \cite%
{ScatteringLambda}, because then we know that it does not contribute to the
product in between $U^{\dag }$ and $U$.

Normally, instead, the fields of the gauge sector are treated as physical
fields (because the gauge symmetry ensures that they mutually compensate
inside the physical quantities). Then the product between $U^{\dag }$ and $U$
is a sum over a complete set of states, which includes the gauge non
invariant ones. Those states are studied by relaxing the initial and final
conditions (\ref{gcond}) on the gauge sector.

\section{Gauge theories on the semi-infinite cylinder}

\label{semiinfinitecylinder}\setcounter{equation}{0}

In this section and the next one we illustrate the general theory in the
cases $\Omega =$ semi-infinite cylinder and $\Omega =$ finite cylinder,
concentrating on the frequencies and the eigenfunctions. We have seen that,
once we have those, we can proceed straightforwardly. We choose the special
gauges $\xi =\lambda $, $\tilde{\xi}=\tilde{\lambda}$, to simplify the
calculations. This allows us to keep one free parameter ($\lambda $) in the
gauge sector and one ($\tilde{\lambda}$) in the purely virtual sector.

We denote the semi-infinite cylinder by $\Omega =S^{1}\times \lbrack -\ell
,\infty )$, while $r$ is the radius of the circle $S^{1}$. Using cylindrical
coordinates $\theta $, $z$, we have%
\begin{equation*}
\bm{A}(t,\theta ,z)=\bm{\hat{\theta}}A_{\theta }(t,\theta ,z)+\bm{\hat{z}}%
A_{z}(t,\theta ,z),\qquad \bm{\nabla }=\frac{\bm{\hat{\theta}}}{r}\frac{%
\partial }{\partial \theta }+\bm{\hat{z}}\frac{\partial }{\partial z}.
\end{equation*}

It is convenient to reach the semi-infinite cylinder from the infinite
cylinder ($\ell =\infty $). We recall that the Lagrangian is (\ref{Ltot})
and the momenta are (\ref{moma}), while the shifted momenta are (\ref{momat}%
). Defining $\Phi =(\phi ,B,A^{0},A_{\theta },A_{z})$, we search for
eigenfunctions of the form%
\begin{equation}
\Phi (t,\theta ,z)=\bar{\Phi}_{0}\mathrm{e}^{i\hat{p}x}\mathrm{e}^{in\theta }%
\mathrm{e}^{-it\hat{\omega}/r},  \label{put}
\end{equation}%
where $\bar{\Phi}_{0}$ denotes a row of constants, while $x=z/r$, $n\in 
\mathbb{Z}$, $\hat{p}$ is a rescaled momentum and $\hat{\omega}$ is a
rescaled frequency. Inserting (\ref{put}) into the field equations derived
from (\ref{Ltot}), the system has solutions when the frequencies are%
\begin{equation*}
\hat{\omega}^{\text{g}}=\frac{1}{\sqrt{\lambda }}\sqrt{n^{2}+\hat{p}^{2}}%
,\qquad \hat{\omega}^{\text{d}}=\frac{1}{\sqrt{\tilde{\lambda}}}\sqrt{n^{2}+%
\hat{p}^{2}},\qquad \hat{\omega}^{\text{ph}}=\sqrt{n^{2}+\hat{p}^{2}}.
\end{equation*}%
Two degeneracies are present, since the gauge-dependent (i.e., $\lambda $%
-dependent) frequencies $\hat{\omega}^{\text{g}}$ and the $\tilde{\lambda}$%
-dependent frequencies $\hat{\omega}^{\text{d}}$ appear twice. Instead, the
physical frequency $\hat{\omega}^{\text{ph}}$ appears once. The independent
solutions for the five components of $\Phi $ are ten: five correspond to the
particles and five correspond to the antiparticles. We do not write their
expressions explicitly. It is sufficient to recall that the most general
solution contains 10 arbitrary constants.

Now we move to the semi-infinite cylinder. Since the $x$ dependence of the
solutions (\ref{put}) is as simple as $\mathrm{e}^{i\hat{p}x}$, they cannot
satisfy the boundary conditions $\Phi (t,\theta ,-\ell )=0$ on $\Omega
=S^{1}\times \lbrack -\ell ,\infty )$, if they are taken separately.
However, if we take linear combinations of functions (\ref{put}) with the
same $\hat{\omega}$, we can impose the conditions $\Phi (t,\theta ,-\ell )=0$
on them. This way, the number of arbitrary coefficients gets reduced to a
half. Ultimately, we obtain five independent solutions, or a solution with
five arbitrary coefficients.

Omitting the overall factor $\mathrm{e}^{in\theta }\mathrm{e}^{-it\hat{\omega%
}/r}$ and the arbitrary constant in front, the physical solution reads%
\begin{eqnarray*}
\phi &=&0,\qquad B=\frac{(\lambda -\tilde{\lambda})\hat{\omega}^{2}}{\tilde{%
\lambda}\hat{p}_{\beta }}\sin \beta ,\qquad A_{0}=-\frac{i\lambda \hat{\omega%
}}{\tilde{\lambda}\hat{p}_{\beta }}\sin \beta , \\
A_{\theta } &=&-\frac{i}{n}\left( \hat{p}_{\alpha }\sin \alpha +\frac{n^{2}}{%
\hat{p}_{\beta }}\sin \beta \right) ,\qquad A_{z}=\cos \alpha -\cos \beta
,\qquad
\end{eqnarray*}%
where%
\begin{equation*}
\alpha =(z+\ell )\frac{\hat{p}_{\alpha }}{r},\qquad \beta =(z+\ell )\frac{%
\hat{p}_{\beta }}{r},\qquad \hat{\omega}^{2}=\hat{p}_{\alpha }^{2}+n^{2}=%
\frac{1}{\lambda }(\hat{p}_{\beta }^{2}+n^{2}).
\end{equation*}%
We see that $\lambda $-dependent contributions are present, but they are
just pure gauge, since the field strength $F=\partial _{z}A_{\theta
}-\partial _{\theta }(A_{z}/r)$ is $\lambda $ independent. It would be
impossible to fulfill the boundary conditions of the semi-infinite cylinder
without a pure gauge part.

To study the limit $\ell \rightarrow \infty $, we multiply the solution by
factors such as $2\mathrm{e}^{\mp i\ell \hat{p}_{\alpha }}$, and drop all
the oscillating terms when $\ell $ gets large. The results are%
\begin{equation*}
\phi ,B,A^{0}\rightarrow 0,\qquad A_{\theta }\rightarrow \mp \frac{\hat{p}%
_{\alpha }}{n}\mathrm{e}^{\pm ix\hat{p}_{\alpha }},\qquad A_{z}\rightarrow 
\mathrm{e}^{\pm ix\hat{p}_{\alpha }},
\end{equation*}%
which coincide with the physical solutions at $\ell =\infty $. If, instead,
we multiply by $2\mathrm{e}^{\mp i\ell \hat{p}_{\beta }}$ and repeat the
same procedure, we obtain an $\ell =\infty $ pure-gauge solution. The other $%
\ell =\infty $ solutions are obtained in similar ways from the general $\ell
<\infty $ solution.

We can identify a solution by the integer $n$, a momentum $\hat{p}$ (e.g., $%
\hat{p}_{\alpha }$ in the example above) and a dispersion relation giving
the frequency $\hat{\omega}$ in terms of $n$ and $\hat{p}$. When we switch
to coherent states, we can label them as%
\begin{equation}
Z_{\hat{p}n}=(z_{\hat{p}n}^{\text{g}},z_{\hat{p}n}^{\text{g\hspace{0.01in}}%
\prime },z_{\hat{p}n}^{\text{d}},z_{\hat{p}n}^{\text{d\hspace{0.01in}}\prime
},z_{\hat{p}n}^{\text{ph}}).  \label{Zn3}
\end{equation}%
The physical solutions correspond to $z_{\hat{p}n}^{\text{ph}}$, and are
quantized as physical particles. We can quantize all the other components of 
$Z_{\hat{p}n}$ as purely virtual particles. This means that we give them
trivial initial and final conditions, and remove the on-shell contributions
due to them, inside the diagrams, perturbatively to all orders, with the
procedures of \cite{diagrammarMio,PVP20}.

The free action is%
\begin{eqnarray}
S_{\text{free}} &=&-i\int \frac{\mathrm{d}\hat{p}}{2\pi }\sum_{n\in \mathbb{Z%
}}(\bar{Z}_{\hat{p}n\text{f}}\Theta _{\hat{p}n}\Omega _{\hat{p}n}Z_{\hat{p}%
n}(t_{\text{f}})+\bar{Z}_{\hat{p}n}(t_{\text{i}})\Theta _{\hat{p}n}\Omega _{%
\hat{p}n}Z_{\hat{p}n\text{i}})  \notag \\
&&+\int_{t_{\text{i}}}^{t_{\text{f}}}\mathrm{d}t\int \frac{\mathrm{d}\hat{p}%
}{2\pi }\sum_{n\in \mathbb{Z}}\left[ i(\bar{Z}_{\hat{p}n}\Theta _{\hat{p}%
n}\Omega _{\hat{p}n}\dot{Z}_{\hat{p}n}-\dot{\bar{Z}}_{\hat{p}n}\Theta _{\hat{%
p}n}\Omega _{\hat{p}n}Z_{\hat{p}n})-2\bar{Z}_{\hat{p}n}\Theta _{\hat{p}%
n}\Omega _{\hat{p}n}^{2}Z_{\hat{p}n}\right] ,\qquad  \label{SZn}
\end{eqnarray}%
where $\Omega _{\hat{p}n}$ is the diagonal matrix of the frequencies, while $%
\Theta _{\hat{p}n}$ is the diagonal matrix of the factors $\mathbb{\tau }%
_{n}=\pm 1$ of (\ref{ortono}).

As we have explained in the previous sections, there is a certain liberty in
choosing the decomposition (\ref{Zn3}), since the only constraints are that: 
$a$) each set is complete for some field $\Phi $ (i.e., it can be used to
expand the field, in order to functionally integrate over it); $b$)
altogether, the eigenfunctions form a complete set for the fields $\Phi $
and the momenta $\bar{\pi}_{\Phi }$; and $c$) the eigenfunctions have the
right limits for $\ell \rightarrow \infty $.

Note that the solutions of the semi-infinite cylinder contain 5 arbitrary
real constants, while those of the infinite cylinder contain twice as many.
They are doubled by the sign choices in the multiplying factors $\mathrm{e}%
^{\pm i\ell \hat{p}_{\alpha }}$, $\mathrm{e}^{\pm i\ell \hat{p}_{\beta }}$,
etc., which are used for the large $\ell $ limit.

It may be puzzling that the number of integration variables of the
functional integral \textquotedblleft doubles\textquotedblright\ in the
limit $\ell \rightarrow \infty $, so to speak. Actually, the number of
variables is infinite, so we cannot really say that it doubles. It is
convenient to explain what happens in detail, since similar instances are
met frequently. Consider the Laplacian on the segment $[0,\ell ]$ with
Dirichlet boundary conditions. We have the eigenfunctions $\sin (\pi nx/\ell
)$, $n\in \mathbb{Z}$, $x\in \lbrack 0,\ell ]$. They \textquotedblleft
double\textquotedblright\ in the limit $\ell \rightarrow \infty $, because,
after centering the segment by means of the shift $x=y+(\ell /2)$, one has
to distinguish the cases $n=$ even and $n=$ odd, which give different
eigenfunctions for $\ell \rightarrow \infty $ (sines and cosines,
respectively). Similarly, $\sin (\ell \hat{p}_{\alpha })=\cos (\ell (\hat{p}%
_{\alpha }-\pi /(2\ell )))$, so the doubling comes from negligible shifts of 
$\hat{p}_{\alpha }$, or $\omega $, which give other eigenfunctions with the
same dispersion relation for $\ell \rightarrow \infty $.

The experimental data we have today, which mainly concern $S$ matrix
amplitudes, are not sufficient to guide us uniquely through the wide freedom
we face when $\tau <\infty $ on a compact $\Omega $. Probably, changing the
basis of physical and purely virtual frequencies in (\ref{Zn3}) is
equivalent to twisting the boundary conditions, or having different
interplays between the experimental setup and the physical process. At any
rate, once we make our choices of initial, final and boundary conditions, as
well as the basis (\ref{Zn3}), everything else is uniquely determined.

\section{Gauge theories on a cylinder}

\label{finitecylinder}\setcounter{equation}{0}

In this section we study gauge theories on a cylinder $\Omega =S^{1}\times
\lbrack -\ell /2,\ell /2]$. We start again from the parametrization (\ref%
{put}) for the solutions of the field equations of the infinite cylinder.
Then we superpose solutions with the same frequency, and impose the boundary
conditions $\Phi (t,\theta ,-\ell /2)=\Phi (t,\theta ,\ell /2)=0$. We find,
as expected, that it is not sufficient to reduce the set of independent
coefficients, as it was for the semi-infinite cylinder, but we must also
discretize the frequencies.

Specifically, we insert%
\begin{equation*}
\Phi (t,\theta ,z)=\tilde{\Phi}(x)\mathrm{e}^{in\theta }\mathrm{e}^{-it\hat{%
\omega}/r}
\end{equation*}%
into the equations, where $x=z/r$, $n\in \mathbb{Z}$, and $\tilde{\Phi}(x)$
are linear combinations of%
\begin{equation*}
\mathrm{e}^{\pm ix\sqrt{\hat{\omega}^{2}-n^{2}}},\qquad \mathrm{e}^{\pm ix%
\sqrt{\tilde{\lambda}\hat{\omega}^{2}-n^{2}}},\qquad \mathrm{e}^{\pm ix\sqrt{%
\lambda \hat{\omega}^{2}-n^{2}}}.
\end{equation*}%
We fix the coefficients of the linear combinations by means of the boundary
conditions, after determining the frequencies $\hat{\omega}$ that admit
nontrivial solutions.

The $\tilde{B}$ equation is independent of the other variables, and just
reads%
\begin{equation}
\tilde{B}^{\prime \prime }=(n^{2}-\lambda \hat{\omega}^{2})\tilde{B},
\label{beq}
\end{equation}%
where the prime denotes the derivative with respect to $x$. Moreover, $%
\tilde{\phi}$ does not enter any equation apart from its own, which reads%
\begin{equation}
\tilde{\phi}^{\prime \prime }=(n^{2}-\lambda \hat{\omega}^{2})\tilde{\phi}%
+\Delta _{\tilde{\phi}},  \label{feq}
\end{equation}%
where $\Delta _{\tilde{\phi}}$ vanishes when all the other fields vanish.
The $\tilde{A}_{0}$ equation depends on $\tilde{A}_{0}$ and $\tilde{B}$,
while the equations of $\tilde{A}_{\theta }$ and $\tilde{A}_{z}$ depend on $%
\tilde{A}_{\theta }$, $\tilde{A}_{z}$ and $\tilde{B}$.

The gauge-dependent frequencies are%
\begin{equation*}
\hat{\omega}_{kn}^{\text{g}}=\frac{1}{\sqrt{\lambda }}\sqrt{k^{2}\frac{\pi
^{2}r^{2}}{\ell ^{2}}+n^{2}},\qquad n\in \mathbb{Z},\qquad k\in \mathbb{N}%
_{+}.
\end{equation*}%
They are associated with two eigenfunctions. One is%
\begin{equation}
\tilde{\phi}_{kn}=\sin \left( k\pi \frac{\hat{z}}{\ell }\right) ,\qquad 
\tilde{B}_{kn}=\tilde{A}_{kn}^{0}=\tilde{A}_{\theta kn}=\tilde{A}_{zkn}=0,
\label{fin}
\end{equation}%
where $\hat{z}=z-(\ell /2)$, and the other one is 
\begin{equation}
\tilde{B}_{kn}=\sin \left( k\pi \frac{\hat{z}}{\ell }\right) ,\qquad \text{%
with nontrivial }\tilde{\phi}_{kn}\text{, }\tilde{A}_{kn}^{0}\text{, }\tilde{%
A}_{\theta kn}\text{ and }\tilde{A}_{zkn}\text{.}  \label{bin}
\end{equation}%
We omit the expressions of the nontrivial fields here, since they are not
crucial for our discussion. The solutions (\ref{bin}) are the only ones with
a nontrivial $\tilde{B}$.

The solutions (\ref{fin}) and (\ref{bin}) are those which, by gauge
invariance, match the eigenfunctions of the ghosts $C$ and $\bar{C}$. Let us
recall that the gauge transformations are $\delta _{\Lambda }\phi =\Lambda
=\theta C$ and $\delta _{\Lambda }\bar{C}=\theta B$. This means that there
must exist $\Phi $ eigenfunctions that are made of $\phi $ only, and match
the $C$ eigenfunctions: these are (\ref{fin}). Moreover, there must exist $%
\Phi $ eigenfunctions where $B$ matches the $\bar{C}$ eigenfunctions: these
are (\ref{bin}). Said in different words, the coherent states that multiply
the solution (\ref{fin}) transform into the $C$ coherent states, while the $%
\bar{C}$ coherent states transform into the coherent states that multiply
the solution (\ref{bin}), as explained in the last part of section \ref%
{cohequad}.

The other frequencies are gauge independent, but depend on $\tilde{\lambda}$%
. Among those, we have the simple frequencies%
\begin{equation}
\hat{\omega}_{kn}^{\text{d}}=\frac{1}{\sqrt{\tilde{\lambda}}}\sqrt{k^{2}%
\frac{\pi ^{2}r^{2}}{\ell ^{2}}+n^{2}},\qquad n\in \mathbb{Z},\qquad k\in 
\mathbb{N}_{+},  \label{omelat}
\end{equation}%
with solutions%
\begin{equation*}
\tilde{\phi}_{kn}=\frac{i\lambda \tilde{A}_{kn}^{0}}{\hat{\omega}_{kn}^{%
\text{d}}(\tilde{\lambda}-\lambda )},\qquad \tilde{A}_{kn}^{0}=\sin \left(
k\pi \frac{\hat{z}}{\ell }\right) ,\qquad \tilde{B}_{kn}=\tilde{A}_{\theta
kn}=\tilde{A}_{zkn}=0.
\end{equation*}

Then we have two other $\tilde{\lambda}$-dependent frequencies, which are
more involved. Their eigenfunctions have $\tilde{B}_{kn}=\tilde{A}%
_{kn}^{0}=0 $, and nontrivial $\tilde{A}_{\theta kn}$, $\tilde{A}_{zkn}$ and 
$\tilde{\phi}_{kn}$. It is straightforward to work them out at $\tilde{%
\lambda}=1$, where the frequencies coincide with (\ref{omelat}). We find $%
A_{\theta }=\sin (k\pi \hat{z}/\ell )$, $A_{z}=0$, and $A_{\theta }=0$, $%
A_{z}=\sin (k\pi \hat{z}/\ell )$.

When $\tilde{\lambda}\neq 1$ it is convenient to expand in powers of $\delta
=(\tilde{\lambda}-1)/2$. The frequencies are%
\begin{eqnarray*}
\hat{\omega}_{kn}^{\text{ph}} &=&\sqrt{k^{2}\frac{\pi ^{2}r^{2}}{\ell ^{2}}+%
\frac{n^{2}}{\tilde{\lambda}(1+\delta ^{2})}}+O(\delta ^{3}), \\
\hat{\omega}_{kn}^{\text{d\hspace{0.01in}}\prime } &=&\sqrt{k^{2}\frac{\pi
^{2}r^{2}}{\tilde{\lambda}\ell ^{2}}+\frac{n^{2}}{1+\delta ^{2}}}+O(\delta
^{3}),
\end{eqnarray*}%
and the solutions read 
\begin{equation*}
A_{\theta }=\sin \left( \frac{k\pi \hat{z}}{\ell }\right) +O(\delta
^{2}),\qquad A_{z}=i\left[ \cos \left( \frac{k\pi \hat{z}}{\ell }\right)
-\cos \left( \frac{k\pi \hat{z}}{\ell }+\frac{nz\delta }{r}\right) \right]
+O(\delta ^{2}),
\end{equation*}%
and%
\begin{equation*}
A_{\theta }=i(-1)^{k}\left[ \cos \left( \frac{k\pi \hat{z}}{\ell }\right)
-\cos \left( \frac{k\pi \hat{z}}{\ell }+(-1)^{k}\frac{nz\delta }{r}\right) %
\right] +O(\delta ^{2}),\quad A_{z}=\sin \left( \frac{k\pi \hat{z}}{\ell }%
\right) +O(\delta ^{2}),
\end{equation*}%
respectively. For the reasons we have explained before, the distinction
between the physical frequencies $\hat{\omega}_{kn}^{\text{ph}}$ and the
purely virtual frequencies $\hat{\omega}_{kn}^{\text{d\hspace{0.01in}}\prime
}$ is to some extent arbitrary.

Once we have the frequencies and the eigenfunctions, we can proceed as in
sections \ref{cohegauge}, \ref{gaugetheories}, \ref{cohequad} and \ref{inter}%
, obtain the coherent-state action (\ref{complaction}), and work out the
evolution operator $U(t_{\text{f}},t_{\text{i}})$ diagrammatically with the
procedure of ref. \cite{MQ}.

\section{Einstein gravity}

\label{gravity}\setcounter{equation}{0}

In this section we study Einstein gravity. The Hilbert-Einstein action%
\begin{equation}
-\frac{1}{16\pi G}\int \mathrm{d}^{4}x\sqrt{-g}R  \label{EE}
\end{equation}%
contains double derivatives of the metric tensor, so it cannot be used as is
to study quantum field theory in a finite interval of time $\tau $ on a
compact manifold $\Omega $. The well-known \textquotedblleft $\Gamma \Gamma $%
\textquotedblright\ action does not have this problem, but differs from (\ref%
{EE}) by a boundary term, which must be treated cautiously, in order to
preserve general covariance. Moreover, in section \ref{cohegauge} we have
emphasized that we need an orthodox symmetry. In particular, the Lagrangian
must be invariant without adding total derivatives, which is not true for
the Hilbert Lagrangian of (\ref{EE}).

The solution of these problems is as follows. First, we perform the purely
virtual extension of ref. \cite{AbsoPhysGrav}, at $\tau =\infty $, $\Omega =%
\mathbb{R}^{3}$. Then, we switch to the invariant metric tensor and
trivialize the symmetry by means of a field redefinition. Third, we add
(invariant) total derivatives and switch to the $\Gamma \Gamma $\ action
(built with the invariant metric tensor). Fourth, we restrict to finite $%
\tau $ and compact $\Omega $ with the procedure of section \ref{cohegauge},
introduce the coherent states, and work out the final action (\ref%
{complaction}). Having trivialized the symmetry, these operations are
invariant.

We begin by recalling the purely virtual extension of gravity at $\tau
=\infty $, $\Omega =\mathbb{R}^{3}$, from \cite{AbsoPhysGrav}. The
gauge-fixed action is\footnote{%
Note some changes of notation with respect to ref. \cite{AbsoPhysGrav}.}%
\begin{equation}
\tilde{S}_{\text{gf}}=-\frac{1}{16\pi G}\int \mathrm{d}^{4}x\sqrt{-g}R+\int
B_{\mu }\left( G^{\mu }(g)-\lambda g^{\mu \nu }B_{\nu }\right) -\int \bar{C}%
_{\mu }\left. \delta _{\xi }\left( G^{\mu }(g)-\lambda g^{\mu \nu }B_{\nu
}\right) \right\vert _{\xi ^{\mu }\rightarrow C^{\mu }},  \label{gf}
\end{equation}%
where $\lambda $ is a gauge-fixing parameter, $G^{\mu }(g)$ is the
gauge-fixing function, $\delta _{\xi }g^{\mu \nu }=\xi ^{\rho }\partial
_{\rho }g^{\mu \nu }-g^{\nu \rho }\partial _{\rho }\xi ^{\mu }-g^{\mu \rho
}\partial _{\rho }\xi ^{\nu }$ is the variation of the inverse metric tensor
under an infinitesimal diffeomorphism $\delta x^{\mu }=-\xi ^{\mu }(x)$, $%
C^{\mu }$ are the Faddeev-Popov ghosts, $\bar{C}_{\mu }$ are the antighosts,
and $B_{\mu }$ are Lagrange multipliers. For example, we can take $G^{\mu
}(g)=\partial _{\nu }g^{\mu \nu }$, or the special gauge of ref. \cite%
{unitarityc}, which is more convenient for various purposes, as it is in
Yang-Mills theory.

Next, we introduce the extra vector $\zeta ^{\mu }(x)$, which by definition
transforms as%
\begin{equation}
\delta _{\xi }\zeta ^{\mu }(x)=\xi ^{\mu }(x-\zeta (x)).  \label{dx}
\end{equation}%
The right-hand side of (\ref{dx}) must be understood as a perturbative
expansion in powers of $\zeta ^{\mu }$. As usual, the Faddeev-Popov ghosts $%
C^{\mu }$ are introduced by writing $\xi ^{\mu }=\theta C^{\mu }(x)$, where $%
\theta $ is a constant anticommuting parameter. Using $\zeta ^{\mu }$, we
can build the invariant metric tensor%
\begin{equation}
g_{\mu \nu \text{d}}(x)=(\delta _{\mu }^{\rho }-\zeta _{,\mu }^{\rho
}(x))(\delta _{\nu }^{\sigma }-\zeta _{,\nu }^{\sigma }(x))g_{\rho \sigma
}(x-\zeta ),  \label{gmunud}
\end{equation}%
where $\zeta _{,\mu }^{\rho }\equiv \partial _{\mu }\zeta ^{\rho }$.

The field $\zeta ^{\mu }$ must be accompanied by anticommuting partners $%
\bar{H}_{\mu }$ and $H^{\mu }$, as well as Lagrange multipliers $E_{\mu }$.
To preserve unitarity, we require $\zeta ^{\mu }$, $\bar{H}_{\mu }$, $H^{\mu
}$ and $E_{\mu }$ to be purely virtual. As in the case of Yang-Mills
theories, the extension amounts to introducing a certain expression in the
functional integral, which is equivalent to \textquotedblleft
1\textquotedblright\ on the $S$ matrix scattering amplitudes, and on the
correlation functions of ordinary (which means $\zeta ^{\mu }$-independent)
insertions of invariant composite fields. However, it allows us to build
new, physical correlation functions, such as those that contain insertions
of the invariant metric tensor (\ref{gmunud}).

Inside the functional integral, the extension is a correction to the action,
which reads%
\begin{equation}
\tilde{S}_{\text{ext}}=\int \mathrm{d}^{4}x\hspace{0.01in}E_{\mu }\left(
V^{\mu }(g,\zeta )-\tilde{\lambda}g_{\text{d}}^{\mu \nu }E_{\nu }\right)
+\int \mathrm{d}^{4}x\hspace{0.01in}\bar{H}_{\mu }\frac{\delta }{\delta
\zeta ^{\rho }}\left( V^{\mu }-\tilde{\lambda}g_{\text{d}}^{\mu \nu }E_{\nu
}\right) H^{\rho },  \label{sext}
\end{equation}%
where $g_{\text{d}}^{\mu \nu }$ is the inverse of $g_{\mu \nu \text{d}}$, $%
V^{\mu }(g,\zeta )$ is an invariant function ($\delta _{\xi }V^{\mu }=0$),
and $\tilde{\lambda}$ is a free parameter. For example, we can take $V^{\mu
}=\partial _{\nu }g_{\text{d}}^{\mu \nu }$, or a mirror of the special gauge.

At this point, we make a change of field variables\footnote{%
Note some different signs with respect to the notation of ref. \cite%
{AbsoPhysGrav}.} 
\begin{equation}
\zeta _{\text{d}}^{\mu }(x)=\zeta ^{\mu }(x+\zeta _{\text{d}}(x)),\qquad C_{%
\text{d}}^{\mu }=(\delta _{\nu }^{\mu }+\zeta _{\text{d},\nu }^{\mu })C^{\nu
},\qquad H^{\mu }=(\delta _{\nu }^{\mu }-\zeta _{,\nu }^{\mu })H_{\text{d}%
}^{\nu },  \label{chv}
\end{equation}%
on the total action $\tilde{S}_{\text{gf}}+\tilde{S}_{\text{ext}}$, to
switch from $g_{\mu \nu }$, $\zeta ^{\mu }$, $C^{\mu }$, $H^{\mu }$ to $%
g_{\mu \nu \text{d}}$, $\zeta _{\text{d}}^{\mu }$, $C_{\text{d}}^{\mu }$, $%
H_{\text{d}}^{\mu }$. We do not change the other fields. This way, we
abandon the original metric tensor $g_{\mu \nu }$ in favor of the invariant
one, $g_{\mu \nu \text{d}}$. Moreover, we trivialize the symmetry, since in
the new variables the transformation of $\zeta _{\text{d}}^{\mu }$ is just $%
\delta \zeta _{\text{d}}^{\mu }=\xi _{\text{d}}^{\mu }\equiv \theta C_{\text{%
d}}^{\mu }$, while $g_{\mu \nu \text{d}}$, $C_{\text{d}}^{\mu }$ and $H_{%
\text{d}}^{\mu }$ are invariant by construction. The trivialized symmetry
thus reads%
\begin{equation}
\delta \zeta _{\text{d}}^{\mu }=\theta C_{\text{d}}^{\mu },\qquad \delta C_{%
\text{d}}^{\mu }=0,\qquad \delta \bar{C}_{\mu }=\theta B_{\mu },\qquad
\delta B_{\mu }=0,  \label{simme}
\end{equation}%
all the other fields being invariant.

Note that%
\begin{equation*}
-\frac{1}{16\pi G}\int \mathrm{d}^{4}x\sqrt{-g}R(g)=-\frac{1}{16\pi G}\int 
\mathrm{d}^{4}x\sqrt{-g_{\text{d}}}R(g_{\text{d}}),
\end{equation*}%
by construction, where $g_{\text{d}}$ inside $\sqrt{-g_{\text{d}}}$ is the
determinant of $g_{\mu \nu \text{d}}$. At this point, we eliminate the
double derivatives by switching to the $\Gamma \Gamma $ action, and restrict
to a finite interval of time $\tau $ and a compact space manifold $\Omega $: 
\begin{equation*}
S_{\Gamma \Gamma }=-\frac{1}{16\pi G}\int_{t_{\text{i}}}^{t_{\text{f}}}%
\mathrm{d}t\int_{\Omega }\mathrm{d}^{3}\bm{x}\sqrt{-g_{\text{d}}}g_{\text{d}%
}^{\mu \nu }(\Gamma _{\mu \lambda \text{d}}^{\alpha }\Gamma _{\nu \alpha 
\text{d}}^{\lambda }-\Gamma _{\mu \nu \text{d}}^{\alpha }\Gamma _{\alpha
\lambda \text{d}}^{\lambda }).
\end{equation*}%
Note that the Lagrangian of this $\Gamma \Gamma $ action, which is built
with the invariant metric tensor, is manifestly invariant, so it satisfies
the identity (\ref{gaugeida}).

The gauge-fixing sector must be rewritten as well, by adding total
derivatives, in order to become invariant at the Lagrangian level. Taking $%
G^{\mu }(g)=\partial _{\nu }g^{\mu \nu }$ for definiteness, we write%
\begin{equation}
S_{\text{gf}}=S_{\Gamma \Gamma }-\int (g^{\mu \nu }-\eta ^{\mu \nu
})(\partial _{\nu }B_{\mu })-\int \lambda B_{\mu }g^{\mu \nu }B_{\nu }+\int %
\left[ \partial _{\nu }\bar{C}_{\mu }+\lambda \bar{C}_{\mu }B_{\nu }\right]
\left. \delta _{\xi }g^{\mu \nu }\right\vert _{\xi ^{\mu }\rightarrow C^{\mu
}},  \label{sgf}
\end{equation}%
where $g_{\mu \nu }$ and $C^{\mu }$ must be understood as functions of $%
\zeta _{\text{d}}^{\mu }$ and $C_{\text{d}}^{\mu }$, according to the change
of variables defined by (\ref{chv}), and $\eta ^{\mu \nu }$ is the
flat-space metric. In (\ref{sgf}) and (\ref{sexta}) below, the integral
symbol stands for the $\mathrm{d}t\hspace{0.01in}\mathrm{d}^{3}\bm{x}$
integral restricted to the interval $\tau $ and the manifold $\Omega $.

The extension (\ref{sext}) is rearranged as%
\begin{equation}
S_{\text{ext}}=\int E_{\mu }\left( \partial _{\nu }g_{\text{d}}^{\mu \nu }-%
\tilde{\lambda}g_{\text{d}}^{\mu \nu }E_{\nu }\right) +\int \left( \partial
_{\nu }\bar{H}_{\mu }+\tilde{\lambda}\bar{H}_{\mu }E_{\nu }\right) \left.
\delta _{\xi }g^{\mu \nu }\right\vert _{g^{\mu \nu }\rightarrow g_{\text{d}%
}^{\mu \nu },\xi ^{\mu }\rightarrow H_{\text{d}}^{\mu }},  \label{sexta}
\end{equation}%
for $V^{\mu }=\partial _{\nu }g_{\text{d}}^{\mu \nu }$, after which we
integrate $E_{\mu }$ away, and proceed as in the case of gauge theories.

We have taken $G^{\mu }(g)=\partial _{\nu }g^{\mu \nu }$ and $V^{\mu
}=\partial _{\nu }g_{\text{d}}^{\mu \nu }$, for concreteness, but it is easy
to adapt the formulas to the special gauge and its mirror, or other choices.

The total action is%
\begin{equation*}
S_{\text{tot}}=S_{\text{gf}}+S_{\text{ext}}
\end{equation*}%
and its symmetry is (\ref{simme}). At this point, we read the Lagrangian $L$
from $S_{\text{tot}}$, and observe that it is orthodoxically symmetric, as
is evident from the expression of (\ref{sgf}), while the Lagrangian of (\ref%
{sexta}) is manifestly invariant. Yet, $L$ contains infinitely many time
derivatives, due to the expansion of expressions like (\ref{dx}) in powers
of $\zeta ^{\mu }$.

The expansion around flat space is defined by writing $g_{\text{d}}^{\mu \nu
}=\eta ^{\mu \nu }+2\kappa h_{\text{d}}^{\mu \nu }$, where $\kappa =$ $\sqrt{%
8\pi G}$ and $G$ is Newton's constant. If we make the replacements%
\begin{eqnarray*}
C_{\text{d}}^{\mu } &\rightarrow &\kappa C_{\text{d}}^{\mu },\qquad B_{\mu
}\rightarrow \kappa ^{-1}B_{\mu },\qquad \bar{C}_{\mu }\rightarrow \kappa
^{-1}\bar{C}_{\mu },\qquad \lambda \rightarrow \lambda \kappa ^{2}, \\
\zeta _{\text{d}}^{\mu } &\rightarrow &\kappa \zeta _{\text{d}}^{\mu
},\qquad H_{\text{d}}^{\mu }\rightarrow \kappa H_{\text{d}}^{\mu },\qquad
E_{\mu }\rightarrow \kappa ^{-1}E_{\mu },\qquad \bar{H}_{\mu }\rightarrow
\kappa ^{-1}\bar{H}_{\mu }\qquad \tilde{\lambda}\rightarrow \tilde{\lambda}%
\kappa ^{2},
\end{eqnarray*}%
the perturbative expansion is the expansion in powers of $\kappa $.

Equations (\ref{chv}) show that $\zeta ^{\mu }\rightarrow \kappa \zeta ^{\mu
}$ plus higher order corrections. The Taylor expansions of arguments such as 
$x^{\mu }-\zeta ^{\mu }$ and $x^{\mu }+\zeta _{\text{d}}^{\mu }$ inside (\ref%
{dx}), (\ref{gmunud}) and (\ref{chv}) raise the powers of $\kappa $ by one
unit for each derivative they generate on the fields. This means that we are
in the situation described in appendix \ref{hdinte}. Applying the
construction of section \ref{cohegauge}, with the rearrangement of appendix %
\ref{hdinte}, we build the correct action (\ref{complaction}) for gravity
restricted to a finite interval of time $\tau $, on a compact space manifold 
$\Omega $. Applying the procedure of \cite{MQ}, we build the evolution
operator $U(t_{\text{f}},t_{\text{i}})$ between arbitrary initial and final
states, with arbitrary boundary conditions, preserving general covariance.

\section{Quantum gravity with purely virtual particles}

\label{PVQG}\setcounter{equation}{0}

The results of the previous section extend to quantum gravity with purely
virtual particles, provided we replace the Hilbert-Einstein action with the
appropriate action.

Since coherent states are \textquotedblleft enemies\textquotedblright\ of
higher derivatives, as we have learned repeatedly, we cannot adopt the
higher-derivative formulation of ref. \cite{LWgrav}, where the Lagrangian
density is made of the Hilbert-Einstein term $R$, plus the cosmological
term, plus the quadratic terms $R^{2}$ and $R_{\mu \nu }R^{\mu \nu }$. We
must start from the two-derivative formulation of ref. \cite{Absograv} at $%
\tau =\infty $, $\Omega =\mathbb{R}^{3}$, which we briefly recall here.

Besides the metric tensor $g_{\mu \nu }$, the theory contains a scalar field 
$\phi $ of mass $m_{\phi }$ (the inflaton) and a spin-2 purely virtual
particle $\chi _{\mu \nu }$ of a certain mass $m_{\chi }$. The action is%
\begin{equation}
S_{\text{QG}}(g,\phi ,\chi ,\Phi )=S_{\text{HE}}(g)+S_{\chi }(g,\chi
)+S_{\phi }(g+\psi ,\phi ),  \label{sew}
\end{equation}%
where 
\begin{equation*}
S_{\text{HE}}(g)=-\frac{r}{16\pi G}\int \sqrt{-g}\left( 2\Lambda
_{C}+R\right) ,\qquad r=\frac{m_{\chi }^{2}}{m_{\phi }^{2}}\frac{3m_{\phi
}^{2}+4\Lambda _{C}}{3m_{\chi }^{2}-2\Lambda _{C}},
\end{equation*}%
is the Hilbert-Einstein action with a cosmological constant $\Lambda _{C}$,%
\begin{equation*}
S_{\phi }(g,\phi )=\frac{3}{4}\left( 1+\frac{4\Lambda _{C}}{3m_{\phi }^{2}}%
\right) \int \sqrt{-g}\left[ D_{\mu }\phi D^{\mu }\phi -\frac{m_{\phi }^{2}}{%
8\pi G}\left( 1-\mathrm{e}^{\phi \sqrt{8\pi G}}\right) ^{2}\right] \qquad
\end{equation*}%
is the inflaton action, and 
\begin{equation*}
S_{\chi }(g,\chi )=S_{\text{HE}}(g+\psi )-S_{\text{HE}}(g)+\int \left[
-2\chi _{\mu \nu }\frac{\delta S_{\text{HE}}(g)}{\delta g_{\mu \nu }}+\frac{%
rm_{\chi }^{2}}{16\pi G}\sqrt{-g}(\chi _{\mu \nu }\chi ^{\mu \nu }-\chi ^{2})%
\right] _{g\rightarrow g+\psi }
\end{equation*}%
is the $\chi _{\mu \nu }$ action, with 
\begin{equation*}
\psi _{\mu \nu }=2\chi _{\mu \nu }+\chi _{\mu \nu }\chi -2\chi _{\mu \rho
}\chi _{\nu }^{\rho },\qquad \chi =\chi _{\mu \nu }g^{\mu \nu }.
\end{equation*}

We gauge-fix (\ref{sew}) as in (\ref{gf}), and make the purely virtual
extension as in (\ref{sext}). Then we switch from the variables $g_{\mu \nu
} $, $\phi $, $\chi _{\mu \nu }$, $\zeta ^{\mu }$, $C^{\mu }$, $H^{\mu }$ to
the variables $g_{\mu \nu \text{d}}$, $\phi _{\text{d}}$, $\chi _{\mu \nu 
\text{d}}$, $\zeta _{\text{d}}^{\mu }$, $C_{\text{d}}^{\mu }$, $H_{\text{d}%
}^{\mu }$, by means of (\ref{gmunud}), (\ref{chv}) and%
\begin{equation*}
\phi _{\text{d}}(x)=\phi (x-\zeta (x)),\qquad \chi _{\mu \nu \text{d}%
}(x)=(\delta _{\mu }^{\rho }-\zeta _{,\mu }^{\rho }(x))(\delta _{\nu
}^{\sigma }-\zeta _{,\nu }^{\sigma }(x))\chi _{\rho \sigma }(x-\zeta ).
\end{equation*}

The action (\ref{sew}) is invariant under the change of variables $g_{\mu
\nu }$, $\phi $, $\chi _{\mu \nu }\rightarrow g_{\mu \nu \text{d}}$, $\phi _{%
\text{d}}$, $\chi _{\mu \nu \text{d}}$, which is just a diffeomorphism. This
means that we can simply view (\ref{sew}) as a function of $g_{\mu \nu \text{%
d}}$, $\phi _{\text{d}}$ and $\chi _{\mu \nu \text{d}}$. Next, we add total
derivatives to eliminate the terms like $\varphi _{1\text{d}}\cdots \varphi
_{n-1\hspace{0.01in}\text{d}}\partial \partial \varphi _{\text{d}n}$ in
favor of terms like $\varphi _{1\text{d}}\cdots \varphi _{n-2\hspace{0.01in}%
\text{d}}\partial \varphi _{n-1\hspace{0.01in}\text{d}}\partial \varphi _{n%
\text{d}}$, in the quadratic sector of the Lagrangian. Moreover, we
rearrange the gauge-fixing part as in (\ref{sgf}) and the purely virtual
extension as in (\ref{sexta}). At that point, we can identify the
eigenfunctions and the coherent states. As far as the interaction sector is
concerned, we rearrange it as explained in appendix \ref{hdinte}. Then we
use the procedure of section \ref{cohegauge} to build the final action (\ref%
{complaction}) for the restriction to finite $\tau $ and compact $\Omega $.
From that point onwards, we can proceed as explained in section \ref%
{cohegauge} and ref. \cite{MQ}, and build the evolution operator $U(t_{\text{%
f}},t_{\text{i}})$ between arbitrary initial and final states, with
arbitrary boundary conditions.

\subsection{Unitarity in the presence of a cosmological constant}

The cosmological constant $\Lambda _{C}$ is nonvanishing, because
renormalization turns it on anyway, even if we start from a vanishing $%
\Lambda _{C}$. A nonzero $\Lambda _{C}$ raises some issues that we must
address.

First of all, flat space is not a solution of the field equations (with $%
\phi =0$, $\chi _{\mu \nu }=0$), so it would be better to formulate
perturbation theory by expanding the metric tensor $g_{\mu \nu }$ around a
de Sitter or anti-de Sitter metric, according to the sign of $\Lambda _{C}$,
rather than the flat-space metric. However, an expansion of that type does
not allow an easy switch to energy/momentum space by means of Fourier
transforms, and makes the calculations of loop diagrams, and the proofs of
general theorems, very hard.

Since the physical results do not depend on the expansion we make, we may
insist on using the expansion around flat space, in spite of its non
standard features. For example, it generates one-leg vertices and a spurious
graviton mass term, which can even be of tachyonic type, depending on the
sign of $\Lambda _{C}$.

Whatever difficulties the expansion may generate, they are of a spurious
nature, which means that they must compensate, and ultimately cancel out. In
this spirit, the expansion around flat space is preferable, because its
unusual features are simpler to deal with.

The other problem concerns the $S$ matrix: we do not know how to define
asymptotic states and $S$ matrix amplitudes on non-flat spacetimes \cite%
{adsscat}. What about unitarity, then?

Although we cannot claim that the $S$ matrix is unitary in a strict sense,
when $\Lambda _{C}\neq 0$, we can still claim that it is unitary up to
effects due to the cosmological constant \cite{ScatteringLambda}. Those
effects are small for all practical purposes: a scattering process should
involve wavelengths as large as the universe to be affected by $\Lambda _{C}$
in a non negligible way.

Besides, now we have a simpler way out. Thanks to the results of this paper
and \cite{MQ}, we are less dependent on the paradigms that have dominated
the scene since the birth of quantum field theory. In particular, we can
study unitarity without being tied to the $S$ matrix, by concentrating on
the evolution operator $U(t_{\text{f}},t_{\text{i}})$.

We have shown that we can build a unitary $U(t_{\text{f}},t_{\text{i}})$
diagrammatically in a finite interval of time $\tau =t_{\text{f}}-t_{\text{i}%
}$, on a compact space manifold $\Omega $, with arbitrary initial and final
states, and arbitrary boundary conditions. The goal has been achieved both
in Einstein gravity (which is not renormalizable, but this does not
jeopardize its perturbative unitarity) and in quantum gravity with purely
virtual particles (which is renormalizable and unitary). In the first case
the cosmological constant can be added with no difficulty, and $U(t_{\text{f}%
},t_{\text{i}})$ remains well-defined and unitary for every $\tau <\infty $.
In the second case, the cosmological constant is already present by default.

This means that the cosmological constant does not have a problem with
unitarity. It does have problems with the very notions of $S$ matrix and
asymptotic states. Given that the difficulties only appear in the $\tau
\rightarrow \infty $ limit, the $\tau <\infty $ formalism we have developed
here might suggest new ways to investigate asymptotic states in gravity with
a cosmological constant.

\section{Conclusions}

\label{conclusions}\setcounter{equation}{0}

When we study gauge theories and gravity on a compact manifold, possibly
with boundary, and on a finite interval of time, we face the nontrivial task
of formulating the initial, final and boundary conditions in invariant ways.
The ordinary gauge potential $A^{\mu }$ and the metric tensor $g_{\mu \nu }$
are not straightforward to handle, in this respect. Nor are the field
strength $F^{\mu \nu a}$, in non-Abelian gauge theories, or the curvature
tensors $R$, $R_{\mu \nu }$, $R_{\mu \nu \rho \sigma }$, in gravity, because
none of them is invariant.

The purely virtual extensions of gauge theories and gravity formulated in
ref.s \cite{AbsoPhys,AbsoPhysGrav} come to the rescue, because they allow us
to define invariant matter and gauge fields $\psi _{\text{d}}$ and $A_{\text{%
d}}^{\mu }$, and an invariant metric tensor $g_{\mu \nu \text{d}}$, without
changing the ordinary physical quantities, such as the $S$ matrix amplitudes
and the correlation functions of nonlinear invariant composite fields, like $%
F_{\mu \nu }^{a}F^{\mu \nu a}$, $\bar{\psi}\psi $, etc. Yet, they allow us
to study new correlation functions, like those of the invariant fields $\psi
_{\text{d}}$, $A_{\text{d}}^{\mu }$ and $g_{\mu \nu \text{d}}$. They also
provide a way of formulating invariant initial, final and boundary
conditions in gauge theories and gravity on a compact manifold $\Omega $, in
a finite interval of time $\tau $.

Switching to the invariant variables $\psi _{\text{d}}$, $A_{\text{d}}^{\mu
} $ and $g_{\mu \nu \text{d}}$, it is also possible to \textquotedblleft
trivialize\textquotedblright\ the symmetries. Then it is relatively
straightforward to organize the action properly, and work out the
eigenfunctions and the frequencies for the expansions of the fields. The
functional integral is defined as the integral on the coefficients of those
expansions. Coherent states are introduced, and the evolution operator $U(t_{%
\text{f}},t_{\text{i}})$ is worked out between arbitrary initial and final
states. The formalism we have developed allows us to calculate $U(t_{\text{f}%
},t_{\text{i}})$ diagrammatically, and perturbatively, for arbitrary
boundary conditions on $\partial \Omega $. In all the operations we make,
the local symmetries are under control, so $U(t_{\text{f}},t_{\text{i}})$ is
gauge invariant and invariant under general coordinate transformations.

We have illustrated the basic properties of the formalism in Yang-Mills
theory on two relatively simple manifolds: the semi-infinite cylinder and
the cylinder.

The limit $\tau \rightarrow \infty $, $\Omega \rightarrow \mathbb{R}^{3}$
(which would give the usual $S$ matrix) is only regular when the
cosmological constant $\Lambda _{C}$ vanishes, due to the problems related
to the definitions of asymptotic states and $S$ matrix amplitudes at $%
\Lambda _{C}\neq 0$. Yet, such problems are not problems of unitarity per
se, because the evolution operator $U(t_{\text{f}},t_{\text{i}})$ of quantum
gravity is unitary for every $\tau <\infty $.

It might be impossible to test the $S$ matrix predictions for a long time,
in quantum gravity. Hopefully, working with $U(t_{\text{f}},t_{\text{i}})$
at finite $\tau $ on a compact $\Omega $ can allow us to explore more
options, and figure out experimental setups that could amplify tiny effects
like those of quantum gravity till they become detectable.

\vskip 1 truecm \noindent {\large \textbf{Acknowledgments}}

\vskip .2 truecm

We are grateful to U. Aglietti for helpful discussions.

\vskip 1truecm

\noindent {\textbf{\huge Appendices}} \renewcommand{\thesection}{%
\Alph{section}} \renewcommand{\theequation}{\thesection.\arabic{equation}} %
\setcounter{section}{0}

\section[Formal properties of delta]{Formal properties of $\delta =\theta
\Delta $}

\label{DeltaProp}\setcounter{equation}{0}

In this appendix we study the key formal properties of the operator $\delta $
of the gauge transformations, and give a very economic proof of theorem (\ref%
{teo}).

Adopting the notation (\ref{smilo}), (\ref{smilad}), (\ref{gaucohe}) of
subsection \ref{gtcohe}, we can write $\delta =\theta \Delta $, 
\begin{equation}
\Delta =\sum_{a}\sum_{n\in \mathcal{\hat{U}}}v_{n}^{aT}\sigma _{1}\frac{%
\delta _{l}}{\delta u_{n}^{a}}+\text{c.c.},\qquad  \label{delta}
\end{equation}%
where \textquotedblleft c.c.\textquotedblright\ denotes the complex
conjugate. It is easy to prove the properties%
\begin{eqnarray}
\Delta ^{2}=0,\qquad &&\Delta u_{n}^{aT}=v_{n}^{aT}\sigma
_{1}+u_{n}^{aT}\sigma _{3}\Delta ,\qquad \Delta v_{n}^{aT}=v_{n}^{aT}\sigma
_{3}\Delta ,  \notag \\
&&\frac{\delta _{l}}{\delta u_{n}^{a}}\Delta =\Delta \hspace{0.01in}\sigma
_{3}\frac{\delta _{l}}{\delta u_{n}^{a}},\qquad \frac{\delta _{l}}{\delta
v_{n}^{a}}\Delta =\Delta \hspace{0.01in}\sigma _{3}\frac{\delta _{l}}{\delta
v_{n}^{a}}+\sigma _{1}\frac{\delta _{l}}{\delta u_{n}^{a}},  \label{prope}
\end{eqnarray}%
where $\Delta $ is meant to act everywhere to its right.

Now we prove theorem (\ref{teo}), stating that a local function $X$ that
solves $\Delta X=0$ can be written as $X=X_{0}+Y$, where $X_{0}=\left.
X\right\vert _{u=v=\bar{u}=\bar{v}=0}$ and $Y$ is a local function.

Define the operators%
\begin{equation*}
\hat{\Delta}=\sum_{a}\sum_{n\in \mathcal{\hat{U}}}u_{n}^{aT}\sigma _{1}\frac{%
\delta _{l}}{\delta v_{n}^{a}}+\text{c.c.},\qquad \mathcal{D}%
=\sum_{a}\sum_{n\in \mathcal{\hat{U}}}\left( u_{n}^{aT}\frac{\delta _{l}}{%
\delta u_{n}^{a}}+v_{n}^{aT}\frac{\delta _{l}}{\delta v_{n}^{a}}\right) +%
\text{c.c.}
\end{equation*}%
Using (\ref{prope}), it is straightforward to prove the identities%
\begin{equation}
\lbrack \mathcal{D},\Delta ]=[\mathcal{D},\hat{\Delta}]=0,\qquad \{\Delta ,%
\hat{\Delta}\}=\mathcal{D}.  \label{dd}
\end{equation}%
The former is a consequence of homogeneity.

Now, decompose $X$ as $X=X_{0}+X^{\prime }$. Clearly, $\Delta X_{0}=\Delta
X^{\prime }=0$. Moreover, $\mathcal{D}^{-1}X^{\prime }$ is well-defined, by
homogeneity. Using (\ref{dd}), we immediately find%
\begin{equation*}
X^{\prime }=\mathcal{DD}^{-1}X^{\prime }=\{\Delta ,\hat{\Delta}\}\mathcal{D}%
^{-1}X^{\prime }=\Delta \hat{\Delta}\mathcal{D}^{-1}X^{\prime }+\hat{\Delta}%
\mathcal{D}^{-1}\Delta X^{\prime }=\Delta Y,\qquad Y=\hat{\Delta}\mathcal{D}%
^{-1}X^{\prime },
\end{equation*}%
which proves the theorem.

Note that we never have to involve, discard, or pay attention to total
derivatives, so the theorem applies to functions, not just functionals.

\section{Higher-derivative interactions}

\label{hdinte}\setcounter{equation}{0}

In this appendix we extend the results of section \ref{cohegauge} to
interaction Lagrangians that contain arbitrarily many derivatives of the
fields, as long as their number grows together with the power of some
coupling. This part is only needed for gravity. We show that we can
rearrange the Lagrangian $\mathcal{L}^{\prime }$ so as to finally have an
action with the form and the properties of (\ref{complaction}).

We assume that the Lagrangian $L(\phi ,\dot{\phi})$ is decomposed as (\ref%
{Lfreeint}), that the symmetry is orthodox and linear, that the quadratic
sector $L_{\text{free}}(\phi ,\dot{\phi})$ has the same structure as in
section \ref{cohegauge} (no more than one derivative on each field, no more
that two derivatives in each term), but we allow $L_{\text{int}}(\phi ,\dot{%
\phi})$ to contain arbitrary monomials $\partial ^{m_{1}}\phi _{1}\cdots
\partial ^{m_{n}}\phi _{n}$ of the fields, differentiated arbitrary numbers
of times $m_{1},\ldots m_{n}$. For definiteness, we assume that $L_{\text{int%
}}(\phi ,\dot{\phi})$ is proportional to some coupling $\lambda $, which we
use to trace the interaction terms. We write them as $\mathcal{O}(\lambda )$%
, or $\mathcal{O}(\lambda ^{n})$, $n>1$, when we mean higher orders.

We proceed as in section \ref{cohegauge} up to the integrated Lagrangian $%
\mathcal{L}^{\prime }$, expressed in terms of coherent states. This means
that:\ we make the shift (\ref{shifto}) with the conditions (\ref{bgaugecond}%
); then we work out the momenta $\tilde{\pi}_{\varphi }^{I}$, make the
redefinition (\ref{pipi2}), and expand $\bar{\tilde{\pi}}_{\varphi
}^{I},\varphi ^{I}$ in coherent states. We obtain the same quadratic part we
had before, then the linear terms due to $\Delta L_{\varphi }^{\prime }$,
plus interactions $\mathcal{L}_{\text{int}}^{\prime }(z,\bar{z})=\mathcal{O}%
(\lambda )$.

Before the expansion in coherent states, we have a wide freedom. For
example, we can change the interaction sector of the Lagrangian by adding
gauge invariant total space derivatives. After the switch from $\bar{\tilde{%
\pi}}_{\varphi }^{I},\varphi ^{I}$ to coherent states, these corrections
give legit vertices. Moreover, the expansion takes care of the space sector,
so we do not need to worry about the space derivatives any longer. What we
have to do, instead, is rearrange the interaction part $\mathcal{L}_{\text{%
int}}^{\prime }(z,\bar{z})$, to remove the time derivatives of $z$ and $\bar{%
z}$, which are still there, and can be arbitrarily many. We achieve this
goal by adding (gauge invariant) total time derivatives to $\mathcal{L}_{%
\text{int}}^{\prime }$.

We can arrange $\mathcal{L}^{\prime }(z,\bar{z})$ into a sum%
\begin{equation}
\mathcal{L}^{\prime }(z,\bar{z})=\mathcal{L}_{\text{free}}^{\prime }(z,\bar{z%
})+\mathcal{L}_{\text{int\hspace{0.01in}0\hspace{0.01in}}}^{\prime }(z,\bar{z%
})+\mathcal{L}_{\text{int\hspace{0.01in}der\hspace{0.01in}}}^{\prime }(z,%
\bar{z}),  \label{leprimo}
\end{equation}%
where $\mathcal{L}_{\text{free}}^{\prime }(z,\bar{z})$ includes the
quadratic terms, as well as the linear terms due to $\Delta L_{\varphi
}^{\prime }$, $\mathcal{L}_{\text{int\hspace{0.01in}0\hspace{0.01in}}%
}^{\prime }(z,\bar{z})=\mathcal{O}(\lambda )$ is free of time derivatives,
while $\mathcal{L}_{\text{int\hspace{0.01in}der\hspace{0.01in}}}^{\prime }(z,%
\bar{z})=\mathcal{O}(\lambda )$ vanishes when all the time derivatives are
set to zero.

We also assume that the each term of $\mathcal{L}_{\text{int\hspace{0.01in}%
der\hspace{0.01in}}}^{\prime }$ has a power of $\lambda $ that is equal to
the number of its time derivatives, at least. We remove $\mathcal{L}_{\text{%
int\hspace{0.01in}der\hspace{0.01in}}}^{\prime }$ iteratively by means of
field redefinitions and dropping gauge invariant total derivatives, without
affecting the symmetry and the other properties of the Lagrangian $\mathcal{L%
}^{\prime }$.

We proceed by induction. We assume that $\mathcal{L}_{\text{int\hspace{0.01in%
}der\hspace{0.01in}}}^{\prime }$ has $N$ powers of $\lambda $ more than one
for each time derivative, and write $\mathcal{L}_{\text{int\hspace{0.01in}der%
\hspace{0.01in}}}^{\prime }=\mathcal{O}(\lambda ^{N})\mathcal{O}(\lambda
\partial _{t})$ to mean this. We give a procedure to rearrange the
Lagrangian so that the new $\mathcal{L}_{\text{int\hspace{0.01in}der\hspace{%
0.01in}}}^{\prime }$ is $\mathcal{O}(\lambda ^{N+1})\mathcal{O}(\lambda
\partial _{t})$. Since we are able to do so for arbitrary $N$, starting from 
$N=0$, we remove $\mathcal{L}_{\text{int\hspace{0.01in}der\hspace{0.01in}}%
}^{\prime }$ entirely.

Replacing the functional derivatives of (\ref{delta}) with ordinary
derivatives, we can write the operator $\Delta $ as 
\begin{equation}
\Delta \equiv \sum_{j=0}^{\infty }\sum_{a}\sum_{n\in \mathcal{\hat{U}}%
}v_{jn}^{aT}\sigma _{1}\frac{\partial _{l}}{\partial u_{jn}^{a}}+\text{c.c}.,
\label{Delta}
\end{equation}%
where $u_{jn}^{a}$ and $v_{jn}^{a}$ denote the $j$-th time derivatives of $%
u_{n}^{a}$ and $v_{n}^{a}$, respectively.

We know that $\mathcal{L}_{\text{int\hspace{0.01in}der\hspace{0.01in}}%
}^{\prime }$ must be gauge invariant by itself ($\Delta \mathcal{L}_{\text{%
int\hspace{0.01in}der\hspace{0.01in}}}^{\prime }=0$), since $\Delta $ does
not mix derivatives and orders of the interactions. Using theorem (\ref{teo}%
), we can write 
\begin{equation*}
\mathcal{L}_{\text{int\hspace{0.01in}der\hspace{0.01in}}}^{\prime
}=X_{0}+\Delta Y,
\end{equation*}%
where $X_{0}$ is a function that depends only on $w_{jn}^{\alpha }$ and $%
\bar{w}_{jn}^{\alpha }$ (the $j$-th time derivatives of $w_{n}^{\alpha }$
and $\bar{w}_{n}^{\alpha }$), and $Y$ is another function.

Since every term $\mathcal{L}_{\text{int\hspace{0.01in}der\hspace{0.01in}}%
}^{\prime }$ must contain time derivatives, $X_{0}$ has the form 
\begin{equation*}
X_{0}=\sum_{j>0}\sum_{\alpha }\sum_{n\in \mathcal{\hat{U}}}w_{jn}^{\alpha
}X_{n}^{\alpha j}+\text{c.c.},
\end{equation*}%
for certain $\Delta $ invariant functions $X_{n}^{\alpha j}$, and their
conjugates. We can write%
\begin{equation*}
X_{0}=\sum_{\alpha }\sum_{n\in \mathcal{\hat{U}}}\dot{w}_{n}^{\alpha
}X_{n}^{\alpha }+X_{0}^{\text{tder}}+\text{c.c.},
\end{equation*}%
where $X_{n}^{\alpha }$ are other $\Delta $ invariant functions, and $X_{0}^{%
\text{tder}}$ are gauge invariant total derivatives. As part of the
rearrangement to get to the correct final action, we drop $X_{0}^{\text{tder}%
}$.

Now we consider $Y$. Since it must contain time derivatives, its form is%
\begin{equation*}
Y=\sum_{j>0}\sum_{n\in \mathcal{\hat{U}}}\left[ \sum_{\alpha }w_{jn}^{\alpha
}Y_{n}^{\alpha j}+\sum_{a}\left(
u_{jn}^{aT}Y_{n+}^{aj}+v_{jn}^{aT}Y_{n-}^{aj}\right) \right] +\text{c.c.}
\end{equation*}%
The only thing that matters is $\Delta Y$, so we can replace $Y$ with 
\begin{equation*}
Y^{\prime }=\sum_{j>0}\sum_{n\in \mathcal{\hat{U}}}\left( \sum_{\alpha
}w_{jn}^{\alpha }Y_{n}^{\alpha j}+\sum_{a}u_{jn}^{aT}\tilde{Y}%
_{n+}^{aj}\right) +\text{c.c.},\qquad \tilde{Y}_{n+}^{aj}=Y_{n+}^{aj}+\sigma
_{1}\sigma _{3}\Delta Y_{n-}^{aj},
\end{equation*}%
since $\Delta Y=\Delta Y^{\prime }$, by the first and third identities of (%
\ref{prope}). Subtracting gauge invariant total derivatives from $\mathcal{L}%
_{\text{int\hspace{0.01in}der\hspace{0.01in}}}^{\prime }$, we rearrange this
expression as 
\begin{equation*}
Y^{\prime }\rightarrow \sum_{n\in \mathcal{\hat{U}}}\left( \sum_{\alpha }%
\dot{w}_{n}^{\alpha }Y_{n}^{\alpha }+\sum_{a}\dot{u}_{n}^{aT}\tilde{Y}%
_{n+}^{a}\right) +\text{c.c.},
\end{equation*}%
for some $Y_{n}^{\alpha }$ and $\tilde{Y}_{n+}^{a}$.

So far, the rearrangement gives%
\begin{equation}
\mathcal{L}_{\text{int\hspace{0.01in}der\hspace{0.01in}}}^{\prime
}\rightarrow \bar{X}\equiv \sum_{n\in \mathcal{\hat{U}}}\left[ \sum_{\alpha }%
\dot{w}_{n}^{\alpha }(X_{n}^{\alpha }+(-1)^{\epsilon _{n}^{\alpha }}\Delta
Y_{n}^{\alpha })+\Delta \sum_{a}\dot{u}_{n}^{aT}\tilde{Y}_{n+}^{a}\right] +%
\text{c.c.,}  \label{rearra}
\end{equation}%
where $\epsilon _{n}^{\alpha }$ is the statistics of $w_{n}^{\alpha }$. Note
that $X_{n}^{\alpha }$, $Y_{n}^{\alpha }$ and $\tilde{Y}_{n+}^{a}$ are $%
\mathcal{O}(\lambda ^{N+1})\mathcal{O}(1)$.

At this point, we remove $\bar{X}$ by means of the field redefinitions, 
\begin{equation}
\bar{w}_{n}^{\alpha }\rightarrow \bar{w}_{n}^{\alpha }-\frac{(-1)^{\epsilon
_{n}^{\alpha }}X_{n}^{\alpha }+\Delta Y_{n}^{\alpha }}{2\mathbb{\tau }%
_{n}^{\alpha }i\omega _{n}^{\alpha }},\qquad \bar{u}_{n}^{a}\rightarrow \bar{%
u}_{n}^{a}-\frac{\sigma _{1}\tilde{Y}_{n+}^{a}}{2\mathbb{\tau }%
_{n}^{a}i\omega _{n}^{a}},\qquad \qquad \bar{v}_{n}^{a}\rightarrow \bar{v}%
_{n}^{a}-\frac{\Delta \tilde{Y}_{n+}^{a}}{2\mathbb{\tau }_{n}^{a}i\omega
_{n}^{a}},  \label{rede}
\end{equation}%
and their conjugates. We show that this operation replaces $\mathcal{L}_{%
\text{int\hspace{0.01in}der\hspace{0.01in}}}^{\prime }$ with higher-order
derivative interactions $\mathcal{O}(\lambda ^{N+1})\mathcal{O}(\lambda
\partial _{t})$, and preserves the key properties of $\mathcal{L}_{\text{free%
}}^{\prime }$ and $\mathcal{L}_{\text{int\hspace{0.01in}0\hspace{0.01in}}%
}^{\prime }$.

When we apply the redefinition (\ref{rede}) to $\mathcal{L}_{\text{int%
\hspace{0.01in}der\hspace{0.01in}}}^{\prime }$ we obtain $\mathcal{O}%
(\lambda ^{N+1})\mathcal{O}(\lambda \partial _{t})$ at least, which go into
the new $\mathcal{L}_{\text{int\hspace{0.01in}der\hspace{0.01in}}}^{\prime }$%
. When we apply (\ref{rede}) to $\mathcal{L}_{\text{free}}^{\prime }$ minus
the universal kinetic terms (\ref{uni1}) and (\ref{uni2}), we obtain: $a$)
interaction terms with no derivatives, which go into the new $\mathcal{L}_{%
\text{int\hspace{0.01in}0\hspace{0.01in}}}^{\prime }$; plus $b$) $\mathcal{O}%
(\lambda ^{N+1})\mathcal{O}(\lambda \partial _{t})$, which go into the new $%
\mathcal{L}_{\text{int\hspace{0.01in}der\hspace{0.01in}}}^{\prime }$. The
same occurs when we apply (\ref{rede}) to $\mathcal{L}_{\text{int\hspace{%
0.01in}0\hspace{0.01in}}}^{\prime }$.

It remains to apply the redefinition (\ref{rede}) to (\ref{uni1}) and (\ref%
{uni2}). The second orders of the Taylor expansions give $\mathcal{O}%
(\lambda ^{N+1})\mathcal{O}(\lambda \partial _{t})$. So, it is sufficient to
focus on the first orders of the Taylor expansions.

From (\ref{uni1}) we get the correction%
\begin{equation*}
-\sum_{\alpha }\sum_{n\in \mathcal{\hat{U}}}\dot{w}_{n}^{\alpha
}(X_{n}^{\alpha }+(-1)^{\epsilon _{n}^{\alpha }}\Delta Y_{n}^{\alpha })+%
\frac{1}{2}\frac{\mathrm{d}}{\mathrm{d}t}\sum_{\alpha }\sum_{n\in \mathcal{%
\hat{U}}}\left[ (-1)^{\epsilon _{n}^{\alpha }}X_{n}^{\alpha }+\Delta
Y_{n}^{\alpha }\right] w_{n}^{\alpha }+\text{c.c.}
\end{equation*}%
The first term subtracts the first one of (\ref{rearra}). The rest is a
gauge invariant total derivative, which we remove.

From (\ref{uni2}) we get the correction 
\begin{equation*}
-\sum_{a}\sum_{n\in \mathcal{\hat{U}}}\Delta (\dot{u}_{n}^{aT}\tilde{Y}%
_{n+}^{a})+\frac{1}{2}\frac{\mathrm{d}}{\mathrm{d}t}\sum_{a}\sum_{n\in 
\mathcal{\hat{U}}}\Delta (u_{n}^{aT}\tilde{Y}_{n+}^{a})+\text{c.c.},
\end{equation*}%
which cancels the rest of (\ref{rearra}), plus gauge invariant total
derivatives.

In the end, we remain with a $\mathcal{L}_{\text{int\hspace{0.01in}der%
\hspace{0.01in}}}^{\prime }$ that is $\mathcal{O}(\lambda ^{N+1})\mathcal{O}%
(\lambda \partial _{t})$. That is to say, we have raised its $\lambda $
power by one unit. Iterating in $N$, we can make $\mathcal{L}_{\text{int%
\hspace{0.01in}der\hspace{0.01in}}}^{\prime }$ disappear entirely.

Summarizing, the effects of the iterated redefinitions (\ref{rede}), the
rearrangements and the droppings of gauge invariant total derivatives in the
interaction sector are: 1) they cancel the term $\mathcal{L}_{\text{int%
\hspace{0.01in}der\hspace{0.01in}}}^{\prime }$; 2) they do not affect the
symmetry transformations (\ref{gaucohe}); this is evident from (\ref{rede}),
using $\delta =\theta \Delta $; 3) they do not affect the universal kinetic
terms (\ref{uni1}) and (\ref{uni2}); 4) they do not affect $\mathcal{L}_{%
\text{free}}^{\prime }$; 5) they do not change the structure of $\mathcal{L}%
_{\text{int\hspace{0.01in}0}}^{\prime }$; 6) they leave the Lagrangian $%
\mathcal{L}^{\prime }$ orthodoxically symmetric. At the end, we have the
correct $\mathcal{L}^{\prime }$: 
\begin{equation}
\mathcal{L}^{\prime }(z,\bar{z})=\mathcal{L}_{\text{free}}^{\prime }(z,\bar{z%
})+\mathcal{L}_{\text{int\hspace{0.01in}0\hspace{0.01in}}}^{\prime }(z,\bar{z%
}).  \label{finalLp}
\end{equation}%
Note that point 6) is tautologically true now: a gauge invariant Lagrangian
of the form (\ref{finalLp}) is necessarily orthodoxically invariant, if the
symmetry is linear, since the universal kinetic terms are invariant by
themselves, and the rest does not contain time derivatives.

The field redefinitions (\ref{rede}) are perturbative, so their Jacobian
determinant is trivial, if we use the analytic or dimensional regularization
techniques \cite{dimreg}.

To get to the action (\ref{action}), we integrate on time, add the usual
endpoint corrections, as in (\ref{action}) and (\ref{complaction}), and we
are done.

\end{document}